\documentclass[%
 reprint,
 showpacs,
 amsmath,amssymb,
 aps,
 prc
]{revtex4-1}

\usepackage{graphicx}
\usepackage{dcolumn}
\usepackage{bm}
\usepackage{csquotes}
\usepackage{diagbox}
\usepackage[caption=false]{subfig}
\usepackage{booktabs}
\usepackage{color}

\newcommand{\bra}[1]{\langle#1|}
\newcommand{\ket}[1]{|#1\rangle}
\renewcommand*{\Im}{\operatorname{Im}} 

\begin{document}

\preprint{APS/123-QED}

\title{Nuclear electromagnetic dipole response with \\ the Self-Consistent Green's Function formalism}

\author{Francesco Raimondi$^{1}$  and Carlo Barbieri$^{1}$}
 
\affiliation{%
 $^1$Department of Physics, University of Surrey, Guildford GU2 7XH, United Kingdom
}%

\date{\today}

\begin{abstract}
\begin{description}
\item[Background] Microscopic calculations of the electromagnetic response of light and medium-mass nuclei are now feasible thanks to the availability of realistic nuclear interactions with accurate saturation and spectroscopic properties, and the  development of large-scale computing methods for many-body physics.
\item[Purpose] To compute isovector dipole electromagnetic (E1) response and related quantities, i.e. integrated dipole cross section and polarizability, and compare with data from photoabsorption and Coulomb excitation experiments. To  investigate the evolution pattern of the E1 response towards the neutron drip line with calculations of  neutron-rich nuclei within a given isotopic chain. 
\item[Methods]  The single-particle propagator is obtained by solving the Dyson equation, where the self-energy includes correlations non-perturbatively through the Algebraic Diagrammatic Construction (ADC) method. The particle-hole ($ph$) polarization propagator is treated in the  Dressed Random Phase Approximation (DRPA), based on an effective correlated propagator that includes some $2p2h$ effects but keeps  the same computation scaling as the standard Hartree-Fock propagator.
\item[Results]  The E1 responses for 
 $^{14,16,22,24}$O, $^{36,40,48,52,54,70}$Ca and $^{68}$Ni have been computed: the presence of a soft dipole mode of excitation for neutron-rich nuclei is found, and there is a fair reproduction of the  low-energy part of the experimental excitation spectrum. This is  reflected in a good agreement with the empirical dipole polarizability values. The impact of different approximations to the correlated propagator used as input in the E1 response calculation is assessed.
 \item[Conclusion] 
 For a realistic interaction that accurately reproduces masses and radii, an effective propagator of mean-field type computed by Self-Consistent Green’s Function provides a good description of the empirical E1 response, especially in the low-energy part of the excitation spectrum and around the Giant Dipole Resonance.  The high-energy part of the spectrum improves and displays an enhancement of the strength when quasiparticle fragmentation is added to the reference propagator. However, this fragmentation (without a proper restoration of dynamical self-consistency) spoils the predictions of the energy centroid of the Giant Dipole Resonance.
\end{description}
\end{abstract}

\pacs{Valid PACS appear here}
\maketitle

\section{\label{intro}Introduction}

The electromagnetic field is an efficient 
tool to investigate the many-body structure of nuclei. The small value of the fine structure constant allows the electromagnetic probe to be treated perturbatively. This simplicity is also reflected in the factorization of the interaction vertex in the electromagnetic and the nuclear density currents, the latter keeping the information on the structure of the initial and final nuclear states~\cite{PaviaBook,HarakehBook}.
Real or virtual photons with different multipolarities can excite bound and resonant nuclear states, probing in this way different internal degrees of freedom of nuclei. For instance, the isovector electromagnetic dipole (E1) field produces the corresponding response, described at the macroscopic level as an out of phase collective oscillation of protons and neutrons~\cite{Teller1948}. This phenomenon, referred to as the Giant Dipole Resonance (GDR), is one of the cornerstones of the nuclear electromagnetic spectroscopy.

With the discovery of the halo nuclei in the eighties~\cite{PhysRevLett.55.2676}, the study of the E1 modes of excitation focused on the low-energy range above the one-nucleon emission threshold: the enhanced E1 response, as soft dipole mode or as Pygmy Dipole Resonance (PDR), is in fact one of the signatures of the neutron-halo and neutron-skin nuclei, respectively~\cite{Aumann2013}.
In general, the dipole response is sensitive to the unbalance of neutron and proton numbers in nuclei, and provides a reaction channel that could be used to constrain the isovector dependence of the nuclear interaction: for instance, models constrained to reproduce the experimental dipole polarizabilities of nuclei are used to predict the corresponding neutron skin thickness~\cite{PhysRevC.92.064304}. In fact, several nuclear quantities related to both infinite nuclear  matter and finite nuclei are correlated to the E1 response and the dipole polarizability: symmetry energy, neutron-skin thickness, charge radii and matter distribution~\cite{PhysRevC.73.044325,PhysRevC.83.034319,PhysRevC.81.051303,PhysRevC.88.024316,ROCAMAZA201896}.
The relation to the equation of state of the neutron-rich matter, which is the macroscopic fermionic system modelling  the inner crust of the neutron stars, ties the dipole response of nuclei to the macroscopic structure of astrophysical objects~\cite{PhysRevC.90.011304,Hagen2016Nature}.

The computation of the E1 excitation spectrum based on a microscopic description of the nuclei  has been achieved first by phenomenological models in the Random Phase Approximation (RPA) and quasiparticle RPA, in both the relativistic and non relativistic frameworks and with and without the effect of the continuum included (see the topical reviews of Refs.~\cite{ROCAMAZA201896,0034-4885-70-5-R02,PhysRevC.94.014304,Co2016}). 
Moreover, several phenomenological approaches going beyond the simple summation of the ring diagrams in the polarization propagator have been formulated: extensions of the RPA with more complex excitation operators~\cite{BRAND19901,Gambacurta2016,PhysRevB.93.165117} also at finite-temperature~\cite{PhysRevLett.121.082501}, the particle-vibration coupling methods such as the Nuclear Field Theory~\cite{1402-4896-91-6-063012} and the Quasiparticle-Phonon Model~\cite{BERTULANI1999139}, the Time-Dependent Density-Functional description of the nuclear dynamics~\cite{RevModPhys.88.045004, LACROIX2004497} and the Extended Theory of Finite Fermi Systems~\cite{KAMERDZHIEV20041}. 

The theoretical approaches starting from realistic nuclear interactions, both conventional and based on chiral effective field theories, are mainly focused to light nuclei, as reviewed in Refs.~\cite{RevModPhys.70.743,0954-3899-41-12-123002}. Recently, the scope of the \textit{ab initio}  many-body methods capable to describe the nuclear excited spectra  has been extended to nuclei with mass number A$>$16. In particular, a series of applications has been put forward within the Coupled Cluster approach combined with the Lorentz Integral Transform (CC-LIT) method, with the computation of the E1 response of several nuclei, from $^4$He   to $^{48}$Ca~\cite{PhysRevC.94.034317,PhysRevC.98.014324}. 

Preliminary calculations of the isovector E1 response and dipole polarizability have also been performed using the Self-Consistent Green's Function (SCGF) approach~\cite{1742-6596-966-1-012015}. Building on this first application, we present in this work  extensive calculations of the E1 response and related quantities of 
medium-mass nuclei, within a formalism in which the particle-hole propagator is treated at RPA level. 
Note that the SCGF formalism is based on expressing the self-energy and particle-hole interaction kernels in terms of skeleton diagrams and fully dressed propagator, rather than mean-field reference states. The self-consistency requirement is a useful feature since it is related to the dynamical fulfilment of conservation laws, however it is not achieved by the dressed RPA (DRPA) many-body truncation used in the present study.
In this work, we exploit the accurate saturation properties of a well-established chiral  two-nucleon (2N) plus three-nucleon (3N) interaction, NNLO$_{\text{sat}}$~\cite{Ekstrom2015}. This chiral interaction is particularly suitable for the computation of quantities related to the nuclear matter distribution and size of the nuclei, since it contains Carbon and Oxygen radii in the pool of fit observables,
and reproduces accurately radii up to the Calcium isotopes~\cite{PhysRevLett.117.052501,Hagen2016Nature}.

Sec.~\ref{sec:formalism} sets out a short review of the  SCGF formalism and the basic equations of the DRPA,  with Sec.~\ref{DIPOLE}  focused on the isovector dipole nuclear response.
After having discussed in Sec.~\ref{sec_conv} the convergence of our calculations with respect to the size and the features of the model space,  we present in the rest of Sec.~\ref{sec:REs}  the results for the E1  photoabsorption cross sections and polarizabilites for several nuclei, from
$^{14}$O to $^{68}$Ni. 
For the closed-subshell nuclei considered below, it is well established that the Dyson formulation of SCGF provides accurate results even when pairing effect are not included explicitly~\cite{PhysRevC.89.061301, PhysRevC.92.014306}.
Different choices of the effective propagators for the DRPA are discussed in Sec.~\ref{other_OpRS}.
Finally, we draw our conclusions in Sec.~\ref{concl}.

\section{\label{sec:formalism}SCGF formalism and E1 nuclear response}

Within the SCGF formalism~\cite{Fetter,Dickhoff2005,Barb2017NLP} the single-particle and the polarization propagators are obtained as solution of the Dyson and Bethe-Salpeter equations, respectively. 
The polarization propagator gives direct access to the nuclear response of an external operator.  Hence, it provides the spectroscopic (overlap functions) and dynamic (energies) information required to compute the nuclear isovector electric dipole response we are interested in.

The spectral information is especially apparent in the Lehmann representation of these propagators. Given the many-body
 Schr\"odinger eigenvalue problem for the $A$- and $A\pm$1-nucleon systems,
\begin{equation}
\label{Schro}
\hat{H} \ \ket{\Psi^{A(\pm 1)}_n} = E_n^{A(\pm 1)} \ \ket{\Psi^{A(\pm 1)}_n} \, ,
\end{equation}
 we consider for the propagation of a single nucleon in the ground state $\ket{\Psi^A_0}$, the one-body Green's function
 \begin{align}
 g_{\alpha \beta}(\omega) ~={}& 
 \sum_{n}  \frac{ 
          \bra{\Psi^A_0}  	a_{\alpha}   \ket{\Psi^{A+1}_n}
          \bra{\Psi^{A+1}_n}  a^{\dagger}_{\beta}  \ket{\Psi^A_0}
              }{\hbar \omega - (E^{A+1}_n - E^A_0)+ \textrm{i} \eta }  \nonumber\\
 +{}& \sum_{k}  \frac{
          \bra{\Psi^A_0}      a^{\dagger}_{\beta}    \ket{\Psi^{A-1}_k}
          \bra{\Psi^{A-1}_k}  a_{\alpha}	    \ket{\Psi^A_0}
             }{\hbar \omega - (E^A_0 - E^{A-1}_k) - \textrm{i} \eta } \; ,
\label{eq:g1}
\end{align}
where the poles give the excitation energies of the \hbox{$A\pm$1-system} with respect to the ground-state energy $E^A_0$,
\begin{align}
\varepsilon_n^{+}\equiv{}&(E^{A+1}_n - E^A_0) \\ 
\varepsilon_k^{-}\equiv{}&(E^A_0 - E^{A-1}_k) \, ,
\end{align}
and the transition amplitudes for the addition and removal of a nucleon are
\begin{align}
\label{tran_ampl}
\langle \Psi_n^{A+1} |a^{\dagger}_{\alpha}|\Psi_0^A \rangle \equiv  {\cal X}^n_{\alpha} \\
~\langle\Psi_k^{A-1}|a_{\alpha}|\Psi_0^A\rangle ~\equiv {\cal Y}^k_{\alpha} \, .
\end{align}

The full expansion of the propagator~(\ref{eq:g1}) in terms of the uncorrelated propagator $ g^{(0)}_{\alpha \beta}(\omega)$ is resummed through the Dyson equation,
\begin{equation}
  \label{eq:Dy}
g_{\alpha\beta}(\omega)=g^{(0)}_{\alpha\beta}(\omega)+ \sum_{\gamma\delta} g^{(0)}_{\alpha\gamma}(\omega)\Sigma_{\gamma\delta}^{\star}(\omega) g_{\delta\beta}(\omega)  \; ,
\end{equation}
which is a nonlinear equation that iterates the irreducible self-energy $\Sigma_{\gamma\delta}^{\star}(\omega)$. The effects of the medium  on the particle propagation are encoded in the self-energy with an organization scheme, the algebraic diagrammatic construction (ADC),  in which the resummation of ring (particle-hole) and ladder (particle-particle and hole-hole) diagrams is performed to all orders~\cite{Barb2017NLP,PhysRevC.97.054308}. 

The Lehmann representation of the polarization propagator is
\begin{align}
 \Pi_{\gamma \delta, \alpha \beta}(\omega) ~={}& 
 \sum_{n_\pi \neq 0}  \frac{ 
          \bra{\Psi^A_0} a^{\dagger}_{\delta}  	a_{\gamma}   \ket{\Psi^{A}_{n_\pi}}
          \bra{\Psi^{A}_{n_\pi}}  a^{\dagger}_{\alpha} 	a_{\beta}   \ket{\Psi^A_0}
              }{\hbar \omega - (E^{A}_{n_\pi} - E^A_0)+ \textrm{i} \eta }  \nonumber\\
 -{}& \sum_{n_\pi \neq 0} \frac{
          \bra{\Psi^A_0}     a^{\dagger}_{\alpha} 	a_{\beta}       \ket{\Psi^{A}_{n_\pi}}
          \bra{\Psi^{A}_{n_\pi}} 	a^{\dagger}_{\delta}  	a_{\gamma}      \ket{\Psi^A_0}
             }{\hbar \omega + (E^{A}_{n_\pi} - E^A_0) -\textrm{i} \eta } \; ,
\label{eq:polariz}
\end{align}
where  $n_{\pi}$ labels the excited states of the $A$-system. In the following, we will use the shorthand notation for the poles,
\begin{equation}
\label{Poles}
\epsilon_{n_{\pi}}^{\pi} \equiv E_{n_\pi}^{A} -E_0^{A} \, ,
\end{equation}
and the residues 
\begin{equation}
\label{Residues}
\mathcal{Z}^{n_{\pi}}_{\alpha \beta} \equiv \bra{\Psi^A_{n_\pi}}  a_{\alpha}^{\dagger} a_{\beta} \ket{\Psi^{A}_0} \, .
\end{equation}
These are  respectively the energies and particle-hole matrix elements between excited states of the A-nucleon system and its ground-state.

The  polarization propagator is solution of the Bethe-Salpeter equation,
\begin{align}
 \Pi_{\gamma \delta, \alpha \beta}(\omega) ~={}&  \Pi^{f}_{\gamma \delta, \alpha \beta}(\omega) \nonumber\\
{}& + \sum_{\mu \rho \nu \sigma} \Pi^{f}_{\gamma \delta, \mu \rho}(\omega)  K^{(ph)}_{\mu \rho,  \nu \sigma}(\omega) \Pi_{\nu \sigma, \alpha \beta}(\omega) \; ,
\label{eq:BetheSal}
\end{align}
where $\Pi^{f}(\omega) $ is the free polarization propagator, and the $ph$ irreducible interaction $K^{(ph)}$ plays for the particle-hole propagator a similar role as that of the self-energy in Eq.~(\ref{eq:Dy}) for the single-particle propagator.

The RPA to Eq.~(\ref{eq:BetheSal}) results from approximating the  $K^{(ph)}$ kernel to first-order, i.e. by using only  the bare interaction vertex.
In standard applications, the associated unperturbed reference propagator is the Hartree-Fock one as required by the Baym-Kadanoff self-consistency approach~\cite{PhysRev.124.287,PhysRev.127.1391}. The RPA can be extended by using the fully correlated single-particle propagator instead of the Hartree-Fock one, yielding the DRPA discussed in next section. 

\subsection{\label{DRPA_sec} Dressed RPA and reduced propagator}

\begin{figure}[t]
{\includegraphics[scale=0.35]{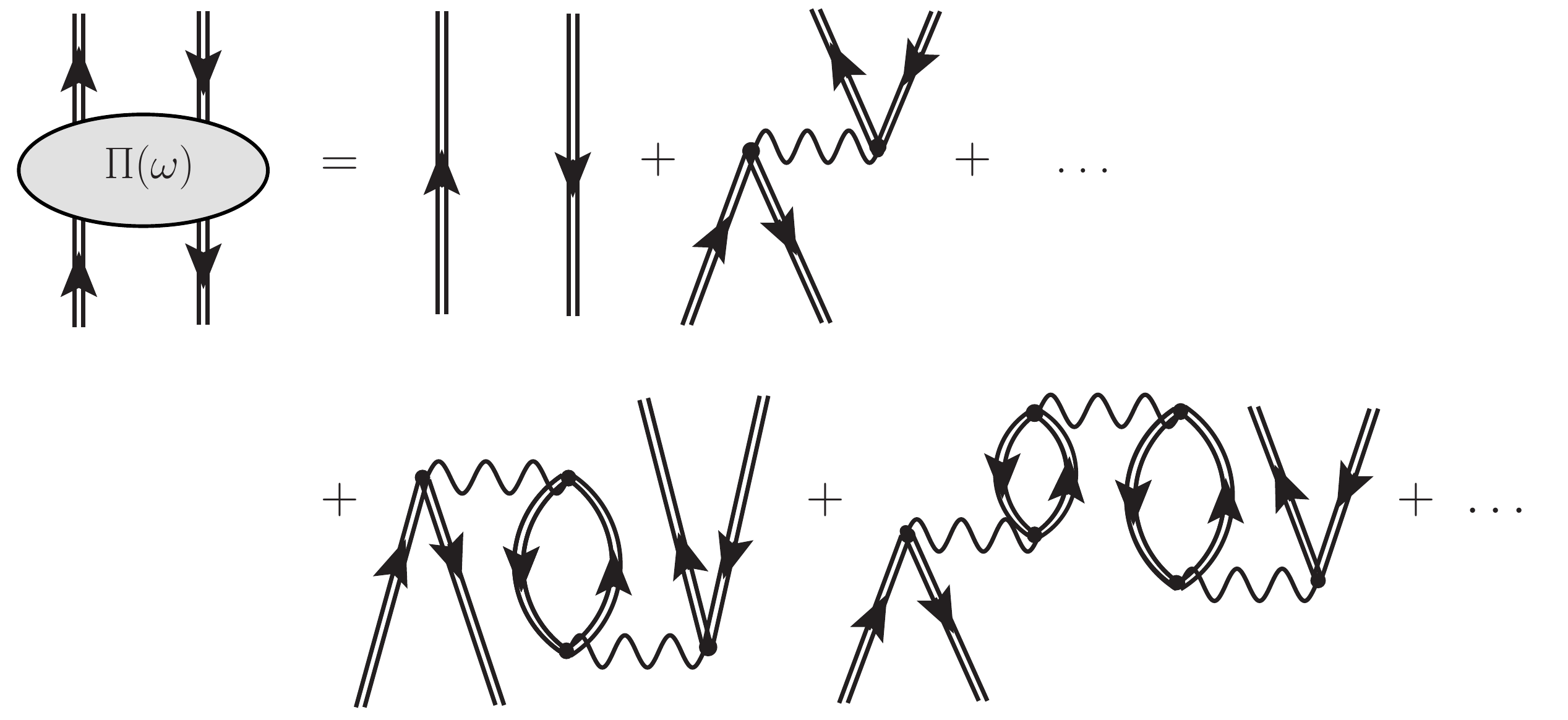}}
  \caption{Espansion of the polarization propagator $\Pi(\omega)$  at the RPA level. Double fermionic lines denote the fully correlated propagators (or OpRS ones), employed in the DRPA.  The expansion truncated at the first row would correspond to the Tamm-Dancoff approximation (TDA).}
  \label{DressedRPA}
\end{figure}

The basic idea of the DRPA is to take into account the fragmentation of the fully correlated propagators in the construction of the free polarization propagator, $\Pi^{f}(\omega)$, as depicted in Fig.~\ref{DressedRPA}. 
The DRPA equation can be cast in the usual matrix form,
\begin{equation}
  \begin{pmatrix}
A &
  B \\
-B^* &
-A^*
   \end{pmatrix}
     \begin{pmatrix}
X \\
Y
   \end{pmatrix} = E      \begin{pmatrix}
X \\
Y
   \end{pmatrix} \; ,
\label{RPA_matrix}
\end{equation}
with the RPA eigenvectors  related to the polarization amplitudes in the following way:
\begin{align}
X_{n k}^{n_{\pi}}=\sum_{\alpha \beta \gamma \delta} \frac{ {\cal X}^n_{\alpha} {\cal Y}^k_{\beta}}{\epsilon_{n_{\pi}}^{\pi} - (\varepsilon_{n}^{+}-\varepsilon_{k}^{-}) } V_{\alpha \gamma \beta  \delta}  \left(\mathcal{Z}^{n_{\pi}}_{\delta \gamma}\right)^{\!*} \\
Y_{n k}^{n_{\pi}}=-\sum_{\alpha \beta \gamma \delta} \frac{ \left({\cal Y}^k_{\alpha} {\cal X}^n_{\beta}\right)^*}{\epsilon_{n_{\pi}}^{\pi} + (\varepsilon_{n}^{+}-\varepsilon_{k}^{-}) } V_{\alpha \gamma \beta  \delta} \left(\mathcal{Z}^{n_{\pi}}_{\delta \gamma}\right)^{\!*} \; .
\end{align}
The submatrices $A$ and $B$ in Eq.~(\ref{RPA_matrix}) are:
\begin{eqnarray}
\label{RPA_matrix_A}
A_{n_1 k_2, n_3 k_4} & = & (\varepsilon_{n_1}^{+}-\varepsilon_{k_2}^{-})\delta_{n_1 n_3} \delta_{k_2 k_4} \nonumber \\ 
&+&  \sum_{\alpha \beta \gamma \delta}  {\cal X}_{\alpha}^{n_1} {\cal Y}_{\beta}^{k_2} V_{\alpha \gamma \beta  \delta} \left( {\cal X}_{\delta}^{n_3} {\cal Y}_{\gamma}^{k_4}  \right)^* \, ,  \\
\label{RPA_matrix_B}
B_{n_1 k_2, n_3 k_4} & =&  \sum_{\alpha \beta \gamma \delta}  {\cal X}_{\alpha}^{n_1} {\cal Y}_{\beta}^{k_2} V_{\alpha \gamma \beta  \delta} {\cal X}_{\gamma}^{n_3} {\cal Y}_{\delta}^{k_4}  \; .
\end{eqnarray}

A study of the $^{16}$O excitation energy spectrum in Ref.~\cite{PhysRevC.68.014311} has shown that the main effects of the fragmentation of the propagator are the screening of the nuclear interaction, with low-lying states pushed at higher energies, and a redistribution of the strength among $ph$ and $2p2h$ phonons considered therein.

Note that  the reference single-particle propagator for the construction of the DRPA matrices, Eqs.~\eqref{RPA_matrix_A} and~\eqref{RPA_matrix_B}, should be the fully correlated propagator. However, the use of dressed propagators increases significantly the requirement in computing resources: the propagator for a typical medium-mass nucleus within an harmonic oscillator model space of 14 major shells contains more than 10$^5$ poles,
which would lead to $ph$ matrices in Eq.~(\ref{RPA_matrix}) which are dense and have dimensions of the order of 10$^{10}$. 
To overcome this limitation, an effective way to include the correlations of the fully dressed propagator has been introduced in Ref.~\cite{Barbieri2009}, with the concept of Optimized Reference State (OpRS) propagator, which we  employ as the reference of our (D)RPA computations. As explained below, the  OpRS propagator includes the relevant many-body correlations while keeping  manageable the computational task at hand. Thus, it  is adopted as the optimal choice for the reference propagator.  The effective OpRS one-body propagator 
\begin{align}
g_{\alpha\beta}^{\rm{OpRS}}(\omega)=\sum_{n \not\in F}\frac{(\psi^n_\alpha)^\ast\psi^n_\beta}{\omega-\varepsilon^{\rm{OpRS}}_n+i\eta}+\sum_{k \in F}\frac{\psi^k_\alpha(\psi^k_\beta)^\ast}{\omega-\varepsilon^{\rm{OpRS}}_k-i\eta} \, ,
\label{g_OpRS}
\end{align}
is obtained by mapping the fully correlated propagator to a simpler one that has a reduced number of poles, for instance one with the same number of poles as the independent particle model (or mean-field) propagator. The effects of the correlations are embedded in the OpRS propagator by requiring that the set of single-particle energies and amplitudes reproduces the first 2$\kappa$ moments of the poles of $g_{\alpha\beta}(\omega)$,
\begin{equation}
  M^p_{\alpha \beta} =
 \sum_n  \frac{ \left( {\cal X}^{n}_{\alpha} \right)^* \;{\cal X}^{n}_{\beta} }
                       {\left[ E_F - \varepsilon^{+}_n \right]^p}  +
 \sum_k \frac{ {\cal Y}^{k}_{\alpha} \; \left( {\cal Y}^{k}_{\beta} \right)^* }
                       {\left[ E_F - \varepsilon^{-}_k \right]^p} \; , 
\label{eq:Mi}
\end{equation}
with $E_F$ being the Fermi energy.
This means that $\varepsilon^{\rm{OpRS}}$ and $\psi_{\alpha}^{n/k}$ are chosen to fulfil the  relations,
\begin{align}
M^{p,\rm{OpRS}}_{\alpha \beta}=M^p_{\alpha \beta} \, ,~ ~  p=0,1,2,\ldots, 2\kappa-1 \, ,
\label{moments}
\end{align}
with integer $\kappa\geq$1.

When the moments of Eq.~(\ref{eq:Mi}) are retained only up to $p=$0 and 1, one obtains an effective OpRS propagator with the same number of poles as the mean-field propagator, corresponding to the single-particle occupancy of the lowest configuration in the independent particle picture. This effective propagator is denoted as $g^{\rm{OpRS}}_{MF}(\omega)$. This is of course the most crude approximation of the dressed propagator:  including higher moments, i.e. for $p>$ 1, allows  for the fragmentation of the single particle strength. The fragmentation becomes denser as higher moments are retained  and the propagator eventually approaches the fully-correlated one.

A feature of Eq.~(\ref{eq:Mi}) is that both particle and hole spectral distributions are mixed together in the same moments. The denominator gives more weight in the sum to those poles closed to $E_F$, hence reproducing at best the correlation effects near the Fermi energy.  Alternatively, one can consider separate moments for the particle and hole distributions using the following definitions:
\begin{equation}
\begin{aligned}
 \widetilde{M}^p_{\alpha \beta} =
 \sum_n  \left( {\cal X}^{n}_{\alpha} \right)^* \;{\cal X}^{n}_{\beta} \left(\varepsilon^{+}_n\right)^p \; ,  \\
 \widetilde{N}^p_{\alpha \beta} = \sum_k  {\cal Y}^{k}_{\alpha} \; \left( {\cal Y}^{k}_{\beta} \right)^* 
                        \left(\varepsilon^{-}_k\right)^p \; , 
\label{eq:Mi_bis}
\end{aligned}
\end{equation}
for $p=0,1,2,\ldots$, which yield an OpRS propagator, denoted in the following as $\widetilde{g}^{\rm{OpRS}}_{p=0,1,2,\ldots}$. Eqs.~(\ref{eq:Mi_bis}) lead to a larger number of poles in $\widetilde{g}^{\rm{OpRS}}_{p}(\omega)$ as compared to Eq.~(\ref{eq:Mi}) but they constrain the particle and hole strengths separately, hereby ensuring that the density profile, total particle number, one-body expectation values and the energy Koltun sum rule of the original propagator are reproduced exactly already for $p\leq$1.

\begin{figure}[t]
{\includegraphics[scale=0.35]{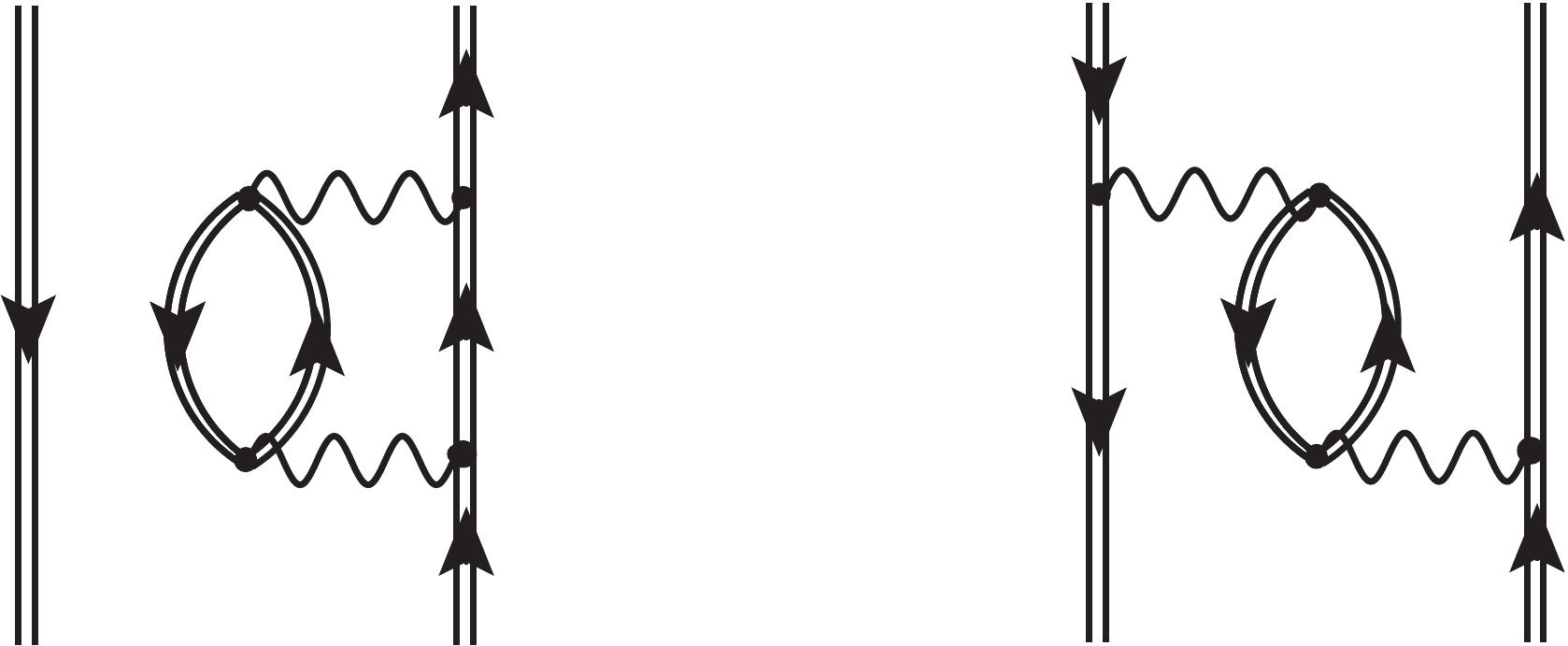}}
  \caption{Example of  diagrams contributing to the $ph$ polarization propagator $\Pi(\omega)$ with $2p2h$ intermediate configurations. Left: non-interacting $1h+2p1h$ terms that contribute to DRPA through the dressing of the reference propagator. Right: interaction among the $ph$ pair mediated by a phonon exchange.}
  \label{bubble}
\end{figure}

It is important to remark that the $g_{\alpha\beta}^{\rm{OpRS}}(\omega)$ propagator, according to the order $p$ of the moments included in the reduction procedure, contains effectively $2p1h$ intermediate state configurations originating from the ADC treatment of the self-energy in the Dyson equation. More specifically, the ADC is implemented by resumming at infinite order the self-energy diagrams topologies at third order,  yielding the ADC(3) scheme~\cite{PhysRevC.97.054308}. For this reason, each particle-hole pair of fermionic lines in the free polarization propagator $\Pi^{f}(\omega)$ can contains in turn $1p-2h1p$, $1h-2p1h$ and $2p1h-2h1p$ intermediate configurations, but in the form of two non-interacting sets of fermionic lines. For instance, both the diagrams in Fig.~\ref{bubble} represent the propagation of a $ph$ pair that includes virtual $2p2h$ intermediate state configurations, but only the diagram 
on the left implicitly contributes to the DRPA since it is composed by a $2p1h$ self-energy  noninteracting with the corresponding hole line.  The diagram on the right side  depicts a particle-hole interaction mediated by a phonon (a bubble).
These bubble diagrams are required to achieve a complete description of $2p2h$ configurations~\cite{BRAND19901,Schirmer1982,doi:10.1063/1.480352}, however they are not included at the DRPA level.  The importance of these terms is also understood by noting that the DRPA could be seen as a hybrid approach because it improves the description of the single-nucleon dynamics by accounting for the fragmentation of its spectral functions, but it continues to approximate the interaction kernel $K^{(ph)}$ at first order. This breaks self-consistency according to the Baym-Kadanoff approach~\cite{PhysRev.124.287,PhysRev.127.1391} so that the fulfilment of fundamental conservation laws is no longer guaranteed. Improving the kernel accordingly, for the ADC(3) polarization, would require a very large number of additional diagrams that also include the bubble exchange of Fig.~\ref{bubble} and other similar terms.  These improvement will be the object of future work. In the present work we will mostly investigate unto which point the fragmentation introduced by DRPA allows to improve the response at large energies, above the giant resonance region.

\subsection{\label{DIPOLE} Isovector Dipole Nuclear Response}
The observables of interest for our purposes are the integrated photoabsorption and Coulomb excitation cross sections, which are computed as
\begin{equation}
\label{CrossSec}
\sigma(E)= 4\pi^2 \alpha  E R(E) \, ,
\end{equation} 
and the dipole polarizability,
\begin{equation}
\label{polariz}
 \alpha_{\text{D}}= 2 \alpha  \int dE \frac{R(E)}{E} \, ,
\end{equation} 
which is the total E1 strength weighted with the inverse of the energy.

Both Eqs.~(\ref{CrossSec}-\ref{polariz}) include the fine-structure constant $\alpha$, and depend on the response $R(E)$ of a nucleus of $Z$ protons and $N$ neutrons  to an  isovector dipole electromagnetic field, with $J^{\pi}T$=1$^{-}$1 quantum numbers,
\begin{equation}
\label{F1}
\hat{\mathcal{Q}}_{1m}^{T=1}=\frac{N}{N+Z} \sum_{p=1}^{Z} r_p Y_{1m} - \frac{Z}{N+Z} \sum_{n=1}^{N} r_n Y_{1m} \, ,
\end{equation}
which is corrected for the center-of-mass displacement, and uses the elementary charge $e$=1. 
The nuclear response contains the matrix element of the field of Eq.~(\ref{F1}) with respect to the correlated excited and ground states,
\begin{equation}
\label{EqMeF}
\bra{\Psi^A_n} \hat{\mathcal{Q}}_{1m}^{T=1}  \ket{\Psi^{A}_0}=  \sum_{\alpha \beta}  \bra{\alpha} \hat{\mathcal{Q}}_{1m}^{T=1} \ket{\beta}  \mathcal{Z}^{n}_{\alpha \beta}  \, ,
\end{equation} 
which is expressed in terms of the single-particle matrix element of the isovector dipole operator and the particle-hole spectroscopic amplitudes of Eq.~(\ref{Residues}). To obtain these amplitudes, we have to solve for the polarization propagator $ \Pi(\omega)$ in the DRPA discussed in Sec.~\ref{DRPA_sec}. 
The corresponding E1 nuclear response is given then by 
\begin{eqnarray}
R(E) &=& - \frac{1}{\pi}  \sum_{\substack{\alpha \beta \\  \gamma \delta}} \bra{\gamma} \hat{\mathcal{Q}}_{1m}^{T=1} \ket{\delta}^* \, \Im{\Pi_{\gamma \delta, \alpha \beta}(E)} \, \bra{\alpha} \hat{\mathcal{Q}}_{1m}^{T=1} \ket{\beta}  \nonumber \\ 
&=&\sum_{n_\pi}  |\bra{\Psi^A_{n_\pi}} \hat{\mathcal{Q}}_{1m}^{T=1}  \ket{\Psi^{A}_0}|^2 \delta(\epsilon_{n_{\pi}}^{\pi}-E) \, .
\label{EqResp}
\end{eqnarray} 
In our discussions below, we will fold the response with a Lorentzian of width $\Gamma$  to smooth the energy dependence,
\begin{equation}
\label{EqResp_folded}
R_{\Gamma}(E) = \sum_{n_\pi}  |\bra{\Psi^A_{n_\pi}} \hat{\mathcal{Q}}_{1m}^{T=1}  \ket{\Psi^{A}_0}|^2 \frac{\Gamma/2\pi}{(\epsilon_{n_{\pi}}^{\pi} - E)^2 + \Gamma^2/4}\, .
\end{equation}

\section{\label{sec:REs}Results}

To calculate the E1 response of a nuclear system to the isovector dipole operator of Eq.~(\ref{F1}), we proceed according to the following steps:
\begin{enumerate}
\item The correlated single-particle propagator for the nucleus of interest is obtained from the Dyson Eq.~(\ref{eq:Dy}), with the self-energy expanded up to ADC(3) in the 2N and 3N interactions, that is by including non-perturbatively correlations  extracted from all Feynman diagrams topologies up third order (see Ref.~\cite{Barb2017NLP,PhysRevC.97.054308} for details). The contributions from 3N forces are included as 2N effective interaction, hence neglecting interaction irreducible 3N terms~\cite{CarAr13,PhysRevC.97.054308};
\item the reduction of the single-particle propagator described in Sec.~\ref{DRPA_sec} is performed and the corresponding effective $g_{\alpha\beta}^{\rm{OpRS}}(\omega)$ is obtained;
\item the quasihole and quasiparticle states of the OpRS propagator are used to build the $ph$ basis spanning the RPA matrices, which are diagonalized to find the solutions of the DRPA Eq.~(\ref{RPA_matrix});
\item the spectroscopic amplitudes obtained from the convergent solution of the Bethe-Salpeter equation are plugged into Eqs.~(\ref{EqMeF}-\ref{EqResp_folded}) to compute the E1 response.
\end{enumerate}

The procedure outlined above has been applied to calculate the E1 photoabsorption cross section and dipole polarizability for light and medium-mass nuclei, from 
$^{14}$O to $^{68}$Ni.

The microscopic Hamiltonian used to compute all the E1 responses in this work is the chiral nuclear interaction NNLO$_{\text{sat}}$~\cite{Ekstrom2015}. The matrix elements of this interaction are computed in Jacobi coordinates and then transformed to an harmonic oscillator (HO) laboratory frame by keeping all matrix elements with $N_1+N_2+N_3\leq$16, where $N_i=2n_i+\ell_i$ is the major oscillator quantum number of nucleon $i$~\cite{Navratil2007tnf}. For our purposes, it is crucial that the spectra of light and medium mass nuclei are  computed with a saturating nuclear Hamiltonian so that both binding energies and radii are reproduced correctly.  Given the established correlation among the matter density distribution and the dipole response in nuclei~\cite{PhysRevC.73.044325}, the NNLO$_{\text{sat}}$ interaction is appropriate for the computation of the E1 response in a microscopic nuclear many-body method.

Concerning the mapping of the fully correlated propagator to the OpRS propagator explained in Sec.~\ref{DRPA_sec}, we have explored different choices in Sec.~\ref{other_OpRS}:  the mean field type $g^{\rm{OpRS}}_{MF}(\omega)$ corresponding to the moments of Eq.~(\ref{eq:Mi}) for $p=$0 and 1, and two $\widetilde{g}^{\rm{OpRS}}_{p}(\omega)$ propagators for  \hbox{$p\leqslant$ 1 or 3}  with the moments defined in Eq.~(\ref{eq:Mi_bis}).

\subsection{\label{sec_conv}Convergence with respect to the model space and the many-body truncation}
In this Section, we discuss the convergence of our calculations with respect to the size of the HO model space. We explore different truncations in terms of the number  of major shells, $N_{\text{max}}$+1, and the HO frequency $\hbar\omega$, considering $^{16}$O as a test case. Convergence should also be gauged with respect to the many-body truncation of the single-particle reference propagator, $g^{\rm{OpRS}}(\omega)$, by starting from the HF approximation and moving to the ADC scheme at second and third order.
It should be clear that, throughout this work, the Bethe-Salpeter equation~\eqref{eq:BetheSal} remains approximated to include only explicit $ph$ (RPA) configurations. 
We also discuss the impact of different choices of the Lorentzian width used to fold the dipole response in Eq.~(\ref{EqResp_folded}).

\begin{figure}
{\includegraphics[width=\columnwidth,keepaspectratio]{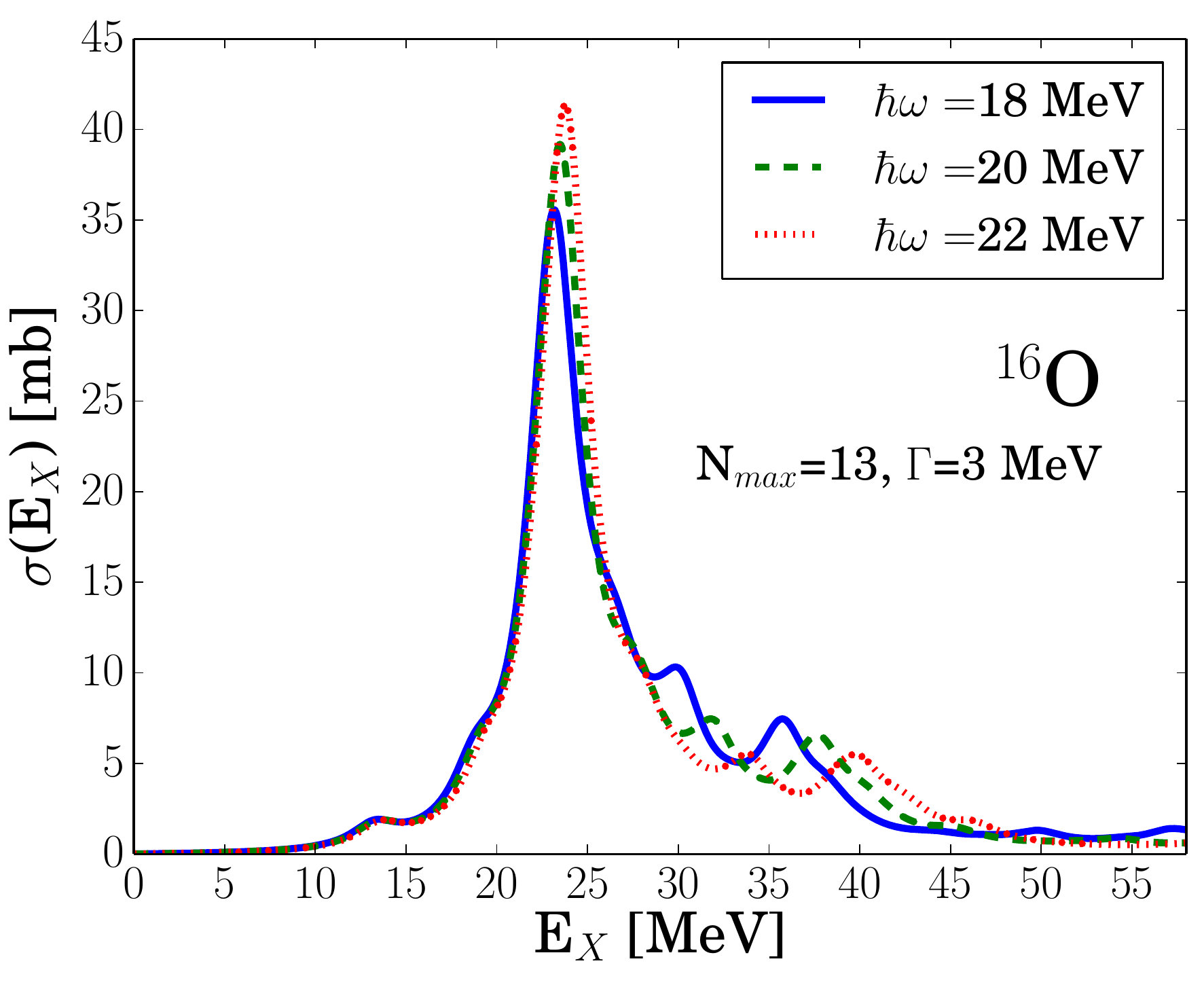}}
\caption{\label{16O_hw} Integrated isovector E1 photoabsorption cross section  of $^{16}$O as function of the excitation energy E$_X$, computed from the NNLO$_{\text{sat}}$ chiral Hamiltonian. The three curves correspond to  $\hbar\omega$=18 (solid line), 20 (dashed line) and 22 (dotted line) MeV, for $N_{\text{max}}$=13 and Lorentzian width $\Gamma$= 3.0 MeV.}
\end{figure}
Fig.~\ref{16O_hw} shows the dependence of photoabsorption cross section of $^{16}$O on the HO frequency $\hbar\omega$ using a model space truncated at  $N_{\text{max}}$=13.  
 The low-energy part of the excitation spectrum up to the position of the giant dipole resonance ($\sim$23.5 MeV) is well converged: this is reflected in the values of the dipole polarizability $ \alpha_{\text{D}}$, displayed in Table~\ref{tab:polariz_conv}. The latter quantity is an inverse energy weighted sum rule (see Eq.~(\ref{polariz})) and it is sensitive to the lower part of the  spectrum. The relative differences between polarizabilities calculated with different nearby values of $\hbar\omega$ is $\sim$1.5\%.
\begin{table}[b]
\caption{\label{tab:polariz_conv} $^{16}$O isovector dipole polarizability $ \alpha_{\text{D}}$ from Eq.~(\ref{polariz}) for  different values of the oscillator parameters $N_{\text{max}}$ and~$\hbar\omega$.}
\begin{ruledtabular}
\begin{tabular}{c||c|c|c}
\backslashbox{$N_{\text{max}}$}{$\hbar\omega$} &  18 MeV & 20 MeV  & 22 MeV                \\
\hline
11  &          0.4997     fm$^3$          &         0.4946     fm$^3$          &            0.4882      fm$^3$           \\
\hline
13 &   0.5026  fm$^3$    &  0.4996   fm$^3$     &   0.4959   fm$^3$       \\
\end{tabular}
\end{ruledtabular}
\end{table}

\begin{figure}
{\includegraphics[width=\columnwidth,keepaspectratio,clip=true,trim=0.0cm 0.0cm 0.0cm 0.05cm]{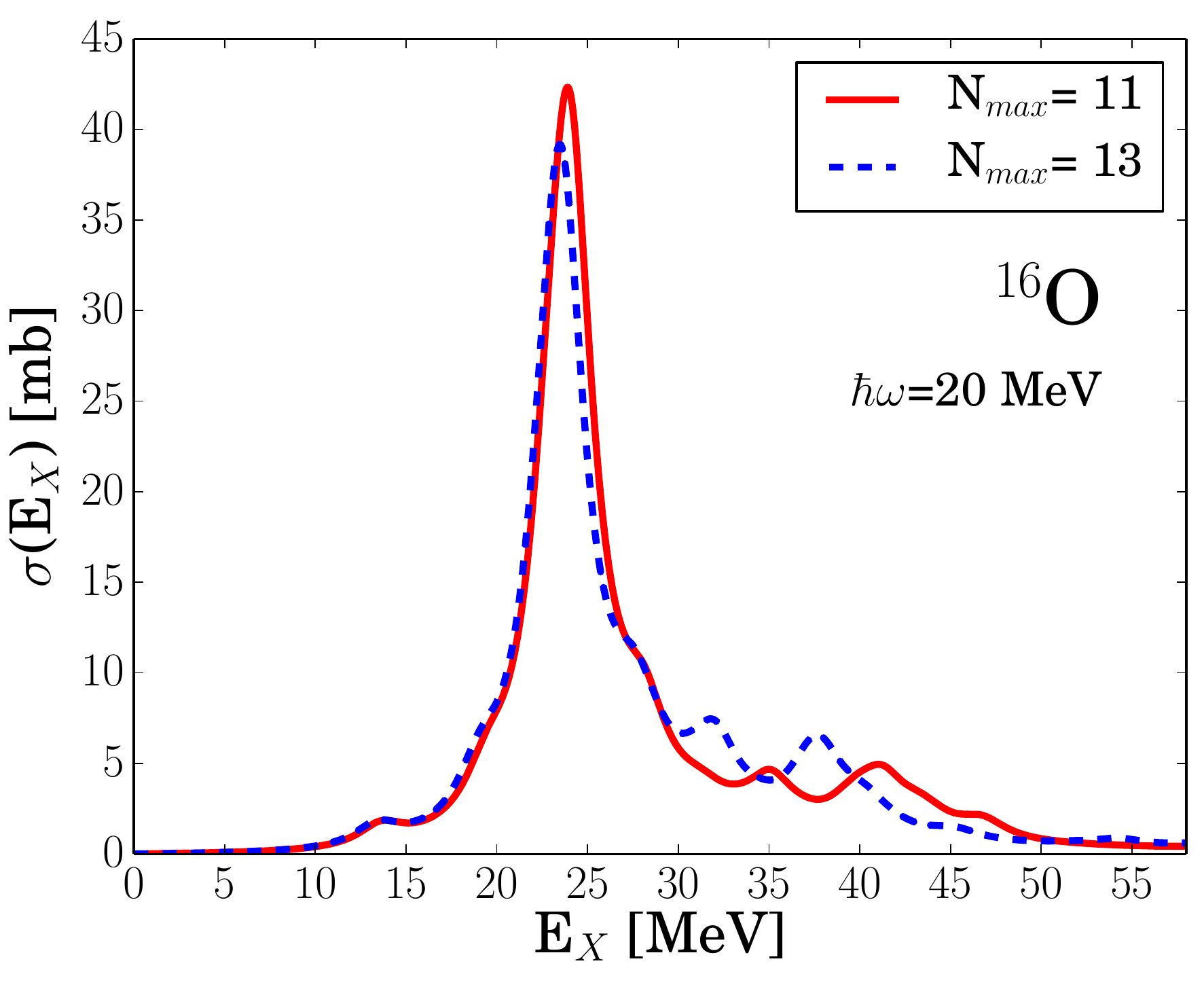}}
\caption{\label{16O_Nmax} Same as in Fig.~\ref{16O_hw} for the HO parameters $N_{\text{max}}$=11 (solid line) and 13 (dashed line) at fixed $\hbar\omega$ = 20 MeV.  }
\end{figure}

The photoabsorption cross section for  $N_{\text{max}}$=11 and 13 are shown in Fig.~\ref{16O_Nmax}. Also in this case the part of the excitation spectrum below the GDR peak is well converged, as confirmed by the two corresponding values of  $ \alpha_{\text{D}}$ in the third column of Table~\ref{tab:polariz_conv} that differ by  $\sim$1\%.

\begin{figure}
{\includegraphics[width=\columnwidth,keepaspectratio,clip=true,trim=0.0cm 0.0cm 0.0cm 0.05cm]{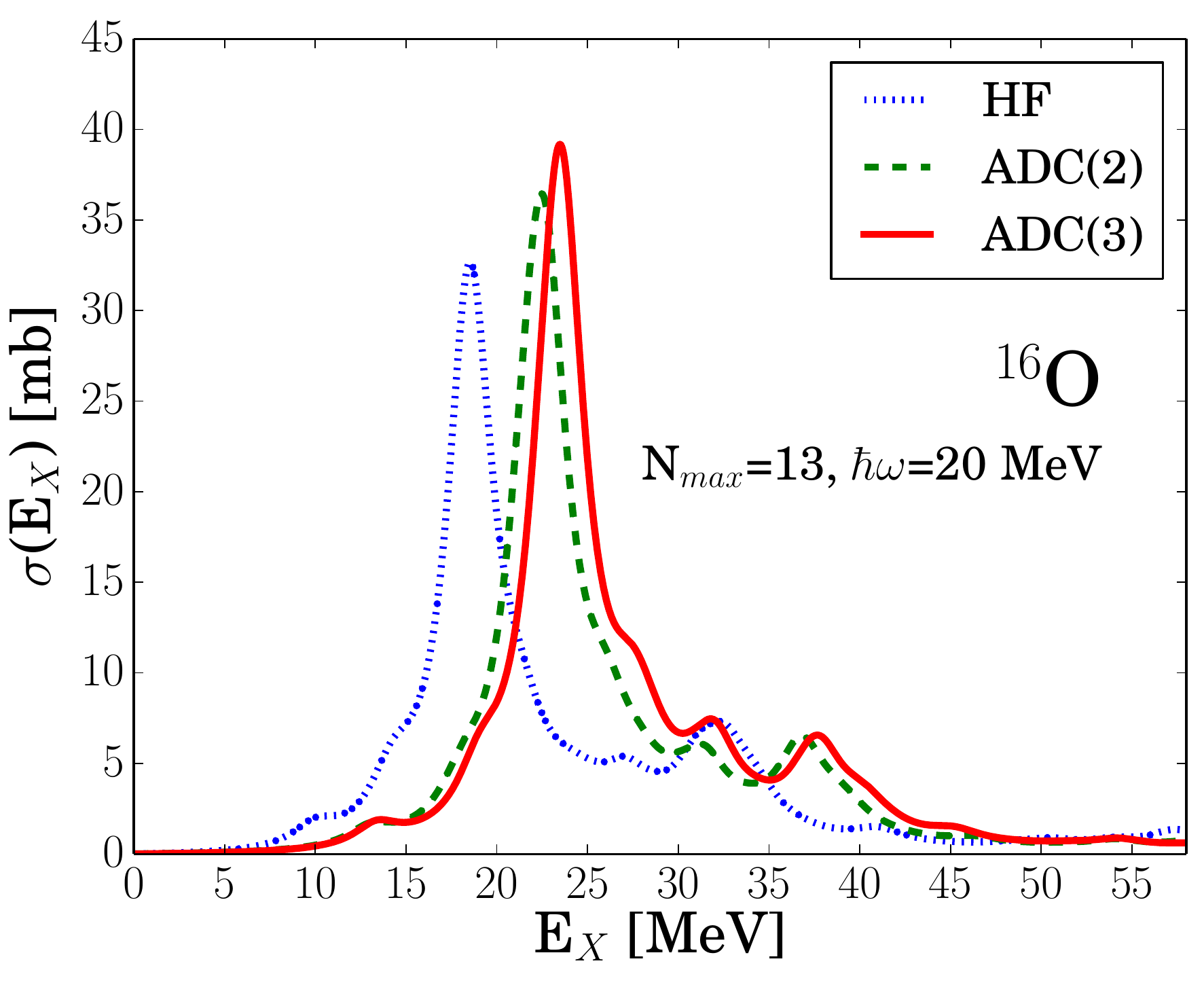}}
\caption{\label{O16_manybody} Same as in Fig.~\ref{16O_hw} but for different many-body truncations of the Dyson propagator, $g(\omega)$, and fixed HO parameters $N_{\text{max}}$=13 and $\hbar\omega$ = 20 MeV. The RPA phonons are built from the uncorrelated HF propagator (dotted line), and from the correlated ones computed using the ADC(2) (dashed line) and ADC(3) (solid lines) truncation schemes.}
\end{figure}

The convergence with respect to the inclusion of correlations in the reference propagator is displayed in Fig.~\ref{O16_manybody}. Here, $g(\omega)$ is computed with different many-body truncations in Eq.~\eqref{eq:Dy} and always reduced to a $g^{\rm{OpRS}}_{MF} (\omega)$ before solving the (D)RPA equations. The comparison between the response calculated from the uncorrelated HF propagator, and the curves obtained from the correlated propagators at ADC(2) and ADC(3) level is instructive: as the structure of the propagator becomes richer, the total strength of the response increases, and the position of the GDR is shifted significantly to higher energy. The stronger effect is seen with the ADC(2), when the lowest order correlations beyond the mean field are introduced in the single-particle propagator. The impact of the ADC(3) is smaller than for the ADC(2) and indicates a saturation of the correlation effects; however, it is decisive for the comparison with the experimental cross section, which we discuss later.   

\begin{figure}
{\includegraphics[width=\columnwidth,keepaspectratio]{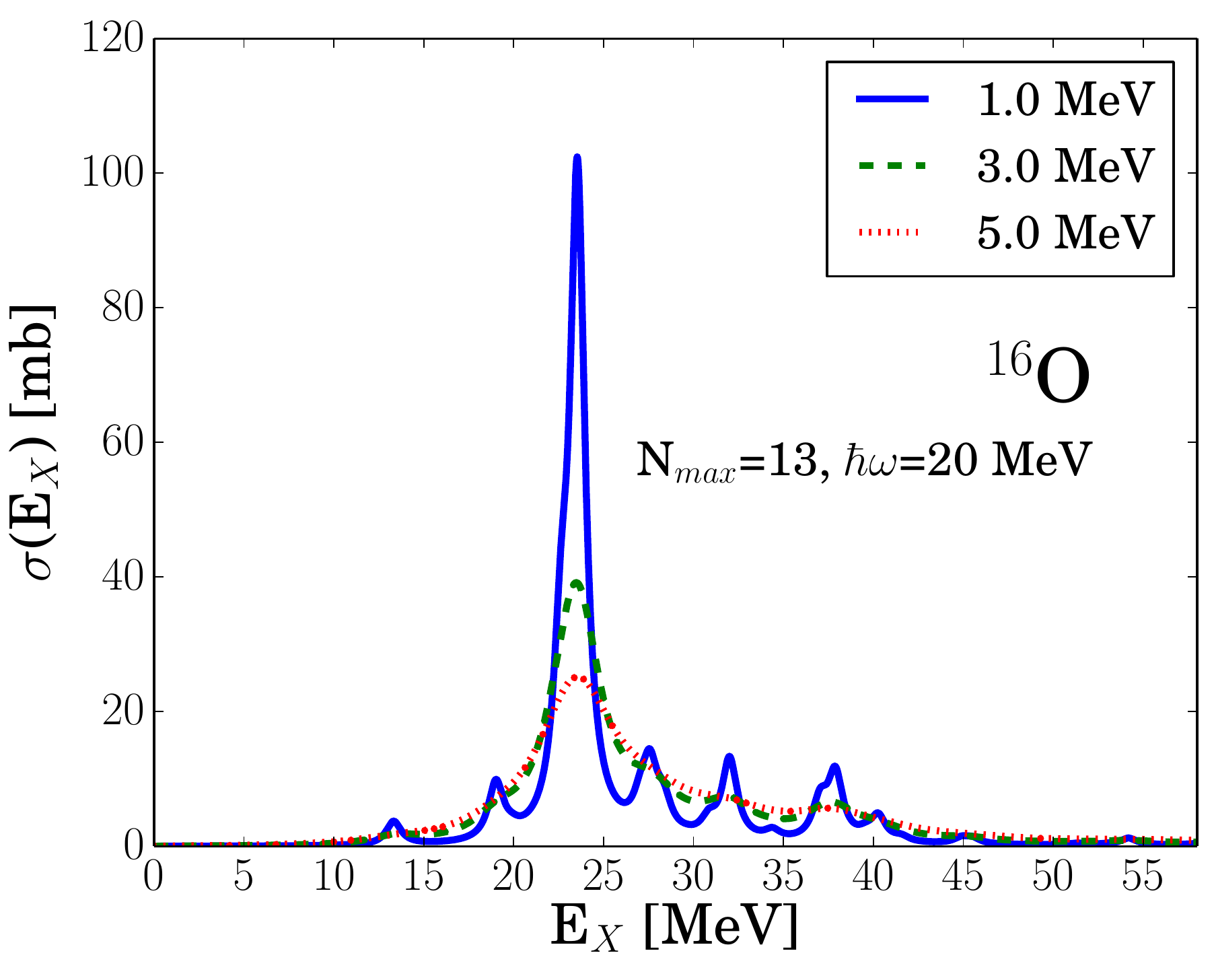}} \\ {\includegraphics[width=\columnwidth,keepaspectratio,clip=true,trim=0.0cm 0.0cm 0.0cm 0.05cm]{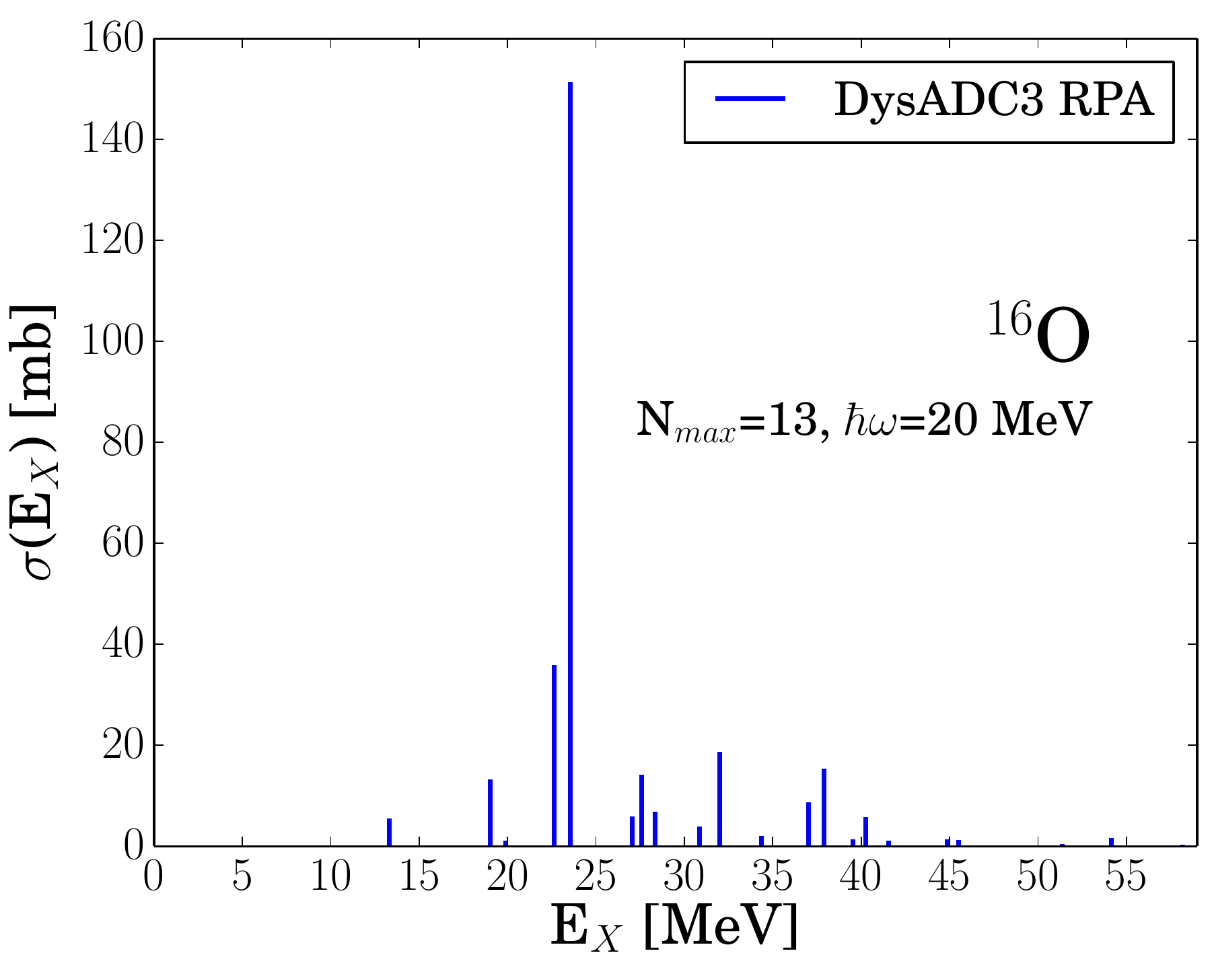}} 
  \caption{  \label{16O_LorWidth} Same as in Fig.~\ref{16O_hw} for the convoluted (upper panel) and the discrete (lower panel) response. 
   The three convoluted curves are obtained by folding the discrete spectrum with Lorentzian widths $\Gamma$ =1.0 (solid line), 3.0 (dashed line) and 5.0 (dotted line) MeV. }
\end{figure}

The convolution of the excitation spectrum with a Lorentzian as in Eq.~(\ref{EqResp_folded}) mimics the continuum effect, by giving a width to the discrete peaks of the response. In principle, provided that all the many-body correlations are included with sufficient precision, the profile of the convoluted spectrum should be insensitive to the chosen width $\Gamma$, for a reasonable range of values. The three curves in the upper panel of Fig.~\ref{16O_LorWidth} show that the position of the GDR peak is not affected by different choices of the width. However, the overall width associated to the resonance depends significantly on $\Gamma$.
 For all the spectra shown in the present work, we consistently choose  $\Gamma$= 3 MeV, corresponding to the value that reproduces the experimental width of the GDR in $^{16}$O (see Fig.~\ref{Oxygens}b).   From the lower panel of Fig.~\ref{16O_LorWidth}, one can see that this value roughly corresponds to the mean separation among the (discrete) dominant RPA eigenstates in the region between 2--35~MeV and therefore it is likely to best describe the experimental width of the giant resonance.   Still, the dependence on $\Gamma$ remains strong at the $ph$-RPA level of the many-body truncation.

\subsection{$^{14,16,22,24}$O}
\begin{figure*}
{\includegraphics[width=\columnwidth,keepaspectratio]{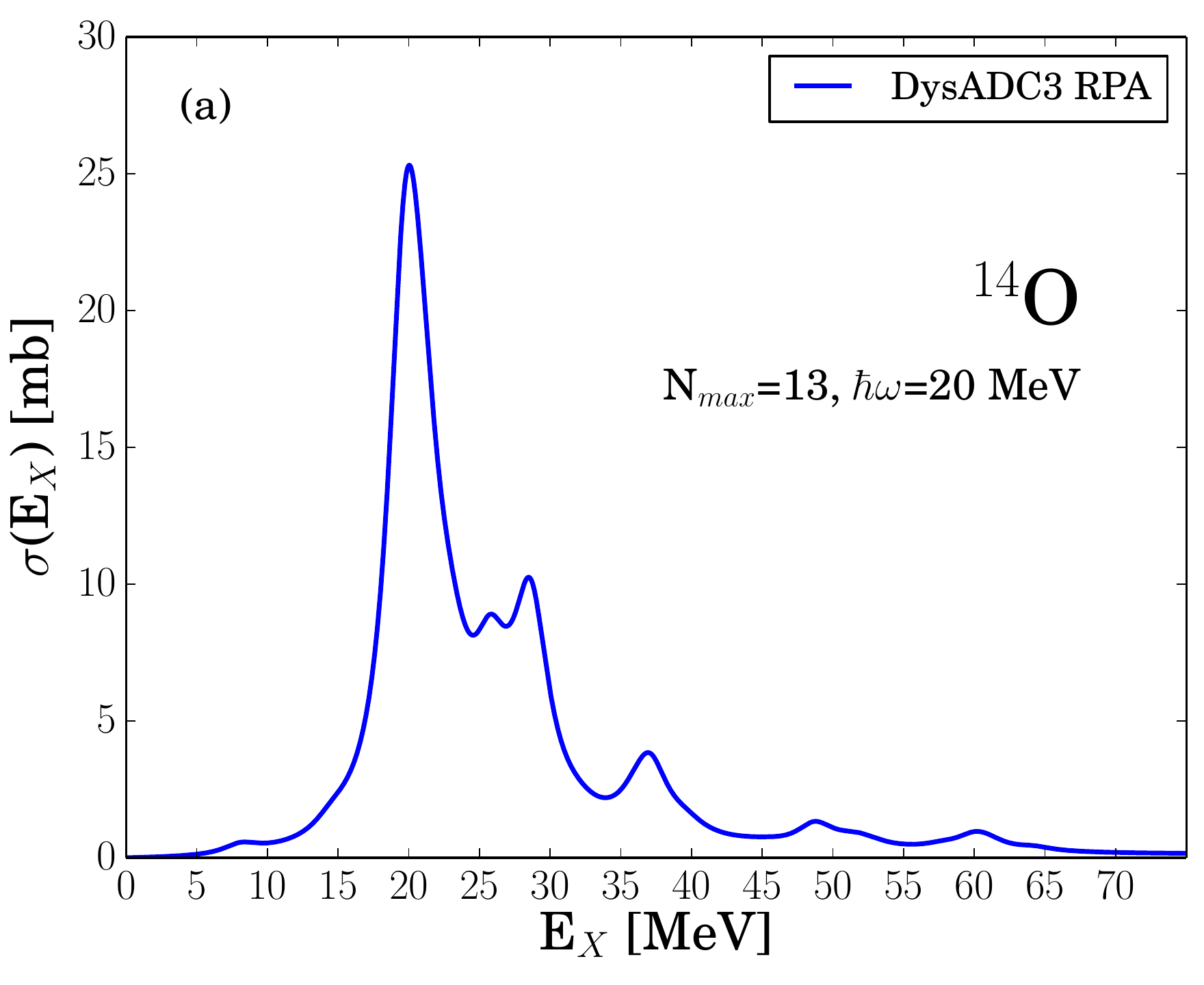}}{\includegraphics[width=\columnwidth,keepaspectratio,clip=true,trim=0.0cm 0.0cm 0.0cm 0.05cm]{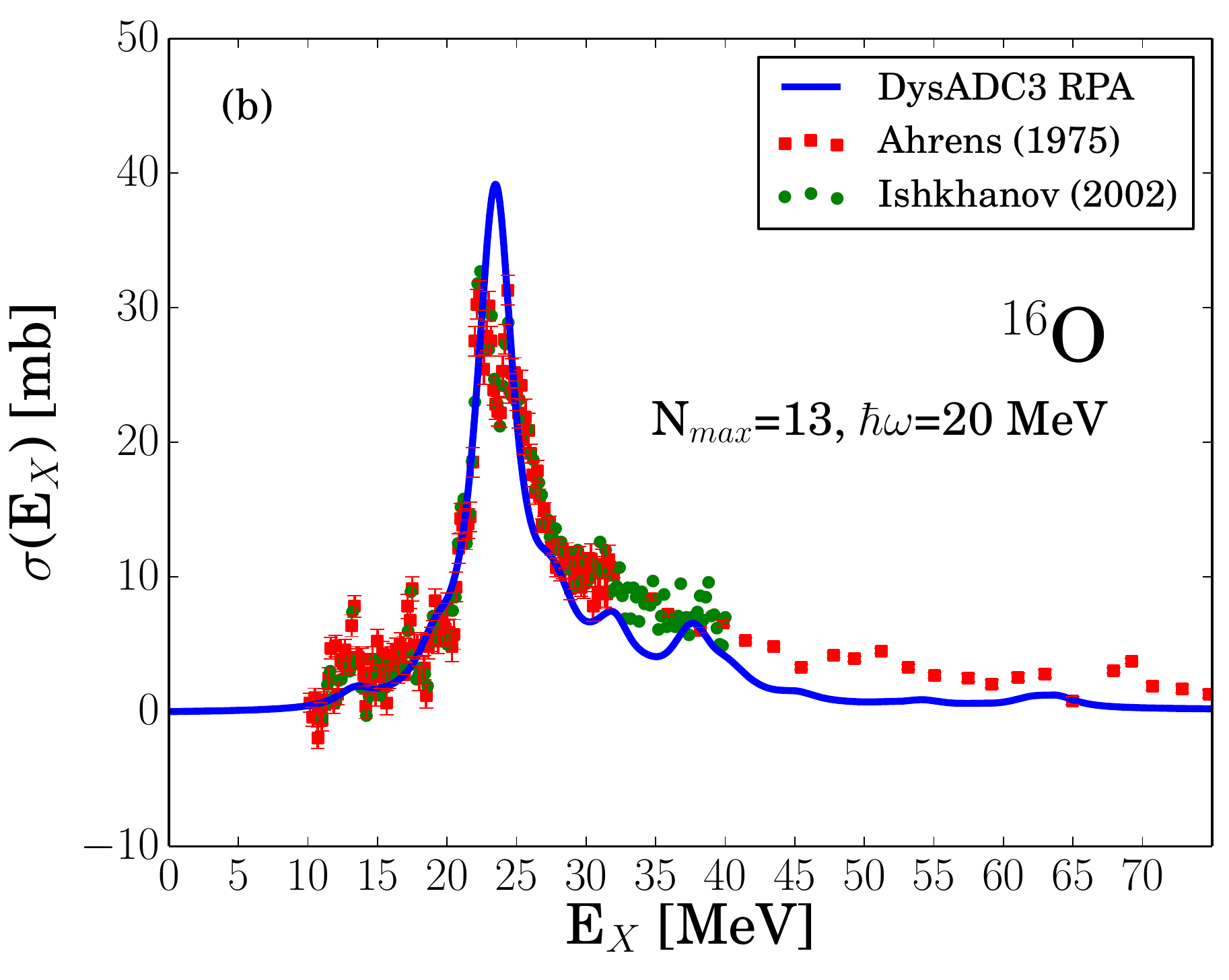}}
{\includegraphics[width=\columnwidth,keepaspectratio]{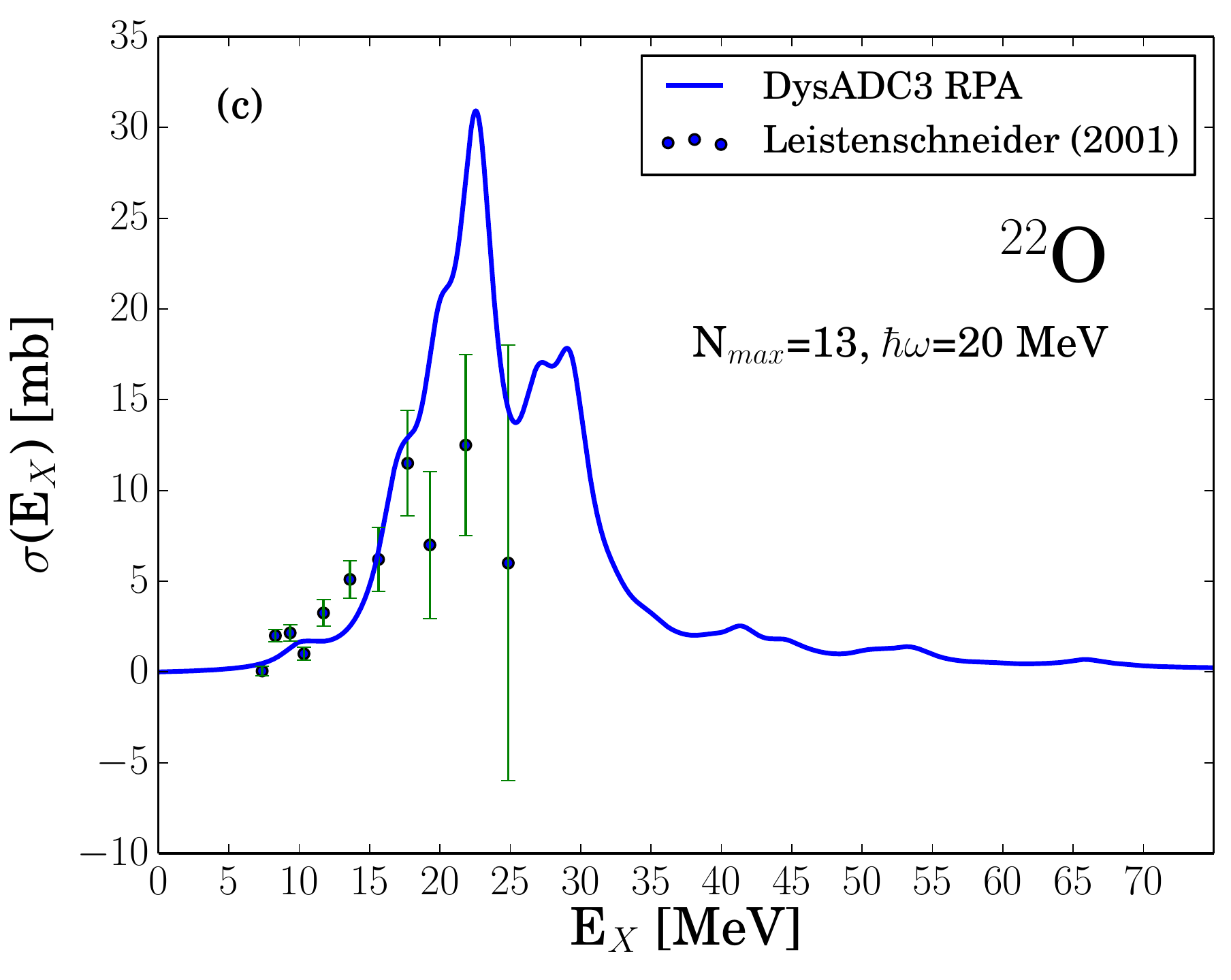}}{\includegraphics[width=\columnwidth,keepaspectratio,clip=true,trim=0.0cm 0.0cm 0.0cm 0.05cm]{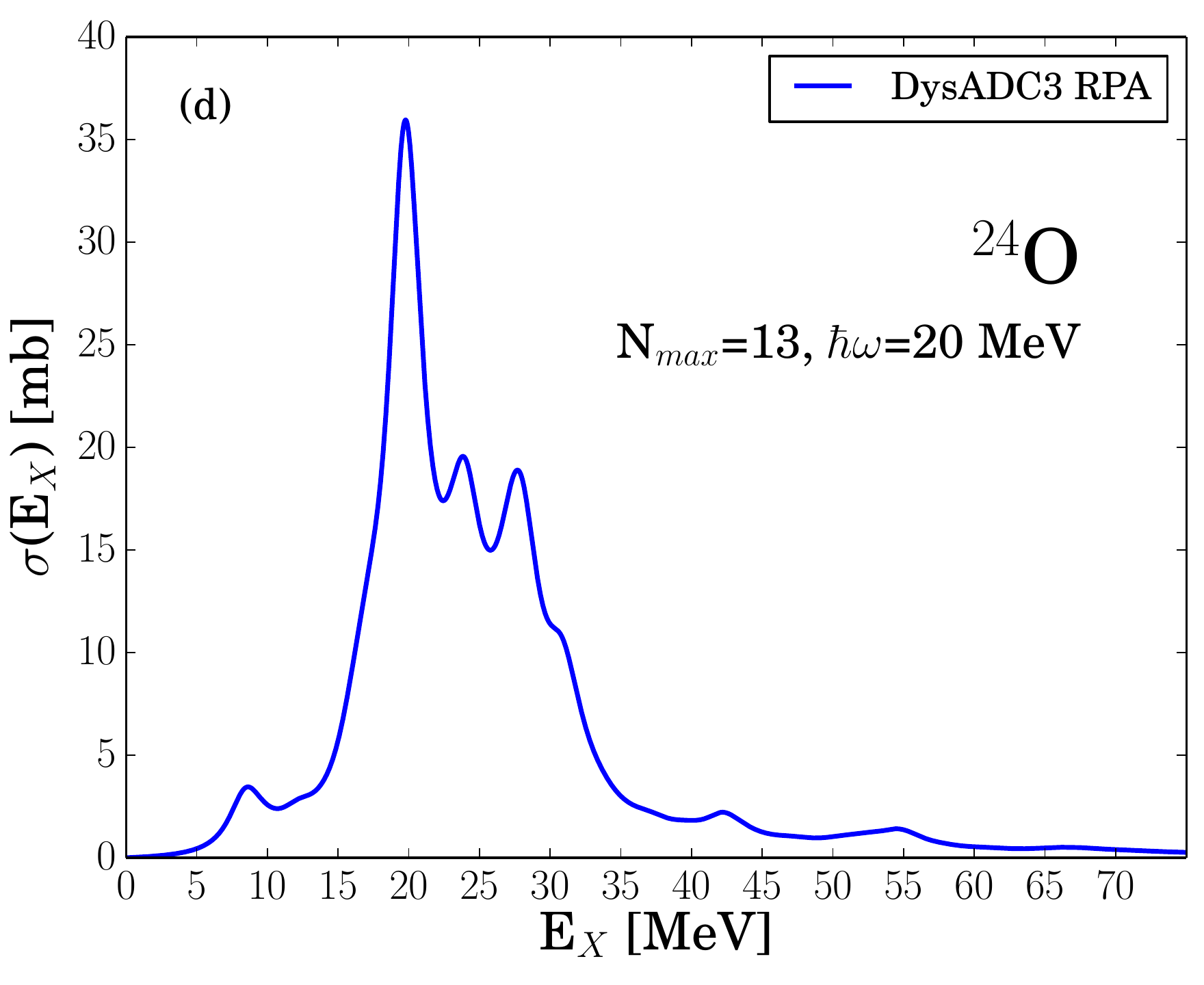}}
\caption{\label{Oxygens} Isovector E1 photoabsorption cross sections of $^{14,16,22,24}$O computed with the NNLO$_{\text{sat}}$ interaction and the SCGF many-body method. The reference $g^{\text{OpRS}}_{MF}(\omega)$ propagator is computed using an ADC(3) self-energy. The curves are obtained by folding the discrete spectra with Lorentzian widths $\Gamma$ = 3.0 MeV. Experimental data for $^{16}$O in panel~(b) are from Ahrens \textit{et al}~\cite{AHRENS1975479} (red squares) and from Ishkhanov \textit{et al}~\cite{ISHKHANOV2002} (green circles); experimental data  for $^{22}$O in panel~(c) are from Leistenschneider \textit{et al}~\cite{PhysRevLett.86.5442}.}
\end{figure*}

Fig.~\ref{Oxygens} shows the isovector E1 photoabsorption cross sections for closed-subshell Oxygen isotopes and using the ADC(3) version of the reference DRPA propagator. The positions of the GDR peak for the four isotopes lie within the range  $\sim$19-24 MeV.  The experimental GDR for $^{16}$O is well reproduced in Fig.~\ref{Oxygens}b, when the spectrum is convoluted with a Lorentzian of $\Gamma$ = 3 MeV which  reproduces the experimental width of the resonance for this measurement. The agreement with the data from Ahrens \textit{et al}~\cite{AHRENS1975479} deteriorates in the high-energy part of the spectrum. The missing computed strength at large energies is due to correlations beyond the $ph$ configurations, that cannot be captured by the simple RPA. 
Table~\ref{tab:polariz_Oxy} compares our results for dipole polarizability, $ \alpha_{\text{D}}$, to the experiment and to the predictions from the CC-LIT many-body method.
 The CC-LIT value for $^{16}$O has been obtained with advanced many-body truncations, including $3p3h$ excitations in both the ground-state and the excited states~\cite{PhysRevC.98.014324}. The 15\% disagreement between the SCGF value and the experiment in $^{16}$O results from the RPA missing both the low-energy strength around 10-15 MeV and strength at large energies above the GDR.
\begin{table}[b]
\caption{\label{tab:polariz_Oxy} $^{16}$O and $^{22}$O isovector dipole polarizabilities $ \alpha_{\text{D}}$  from Eq.~(\ref{polariz}) compared to the CC-LIT calculations of Refs.~\cite{PhysRevC.94.034317,PhysRevC.98.014324}  and to those extracted from the experimental spectra~\cite{AHRENS1975479,PhysRevLett.86.5442}. The $^{22}$O $ \alpha_{\text{D}}$ values are obtained by integrating over the entire energy range (upper values) or up to 3~MeV above the continuum threshold (lower values). }
\begin{ruledtabular}
\begin{tabular}{c||c|c|c}
Nucleus &  SCGF & CC-LIT  & Exp                \\
\hline
$^{16}$O                               &          0.4996   fm$^3$           &         0.528(21)     fm$^3$          &            0.585(9)   fm$^3$              \\
$^{22}$O                              &   0.724  fm$^3$                     &  0.86(4) fm$^3$            &   0.43(4) fm$^3$ \\
$^{22}$O~($E_x<$ 3~MeV)   &    0.05 fm$^3$                      &  0.05(1) fm$^3$                &   0.07(2) fm$^3$       \\
\end{tabular}
\end{ruledtabular}
\end{table}

For the neutron-rich isotopes $^{22}$O and $^{24}$O in Figs.~\ref{Oxygens}c and~\ref{Oxygens}d respectively, part of the dipole strength is redistributed at lower energies: the presence of a soft dipole mode at $\sim$10 MeV can be related to the concept of PDR. In particular,  the cross section for $^{22}$O shows a peak of small strength compatible with the resonance measured at 9.36($\pm$ 0.45) MeV by Leistenschneider \textit{et al}~\cite{PhysRevLett.86.5442}. This implies an enhancement of the dipole polarizability, as expected for neutron-rich nuclei and confirmed by the theoretical calculations displayed in Table~\ref{tab:polariz_Oxy}. The experimental spectrum in Ref.~\cite{PhysRevLett.86.5442} is affected by significant uncertainties in the GDR region, due to the presence of other reaction channels not excluded in the measurement. For this reason, we follow the analysis of Ref.~\cite{PhysRevC.94.034317} and compute the dipole polarizabilities also in a restricted energy range. Table~\ref{tab:polariz_Oxy} shows the results obtained including excitations up to 3 MeV above the continuum threshold: in our discrete response, this range includes two peaks accounting for a fraction of \hbox{$ \alpha_{\text{D}}$= 0.05 fm$^3$}, in agreement with both the experiment and CC-LIT results.

\subsection{$^{36,40,48,52,54,70}$Ca}
\begin{figure*}
{\includegraphics[width=\columnwidth,keepaspectratio]{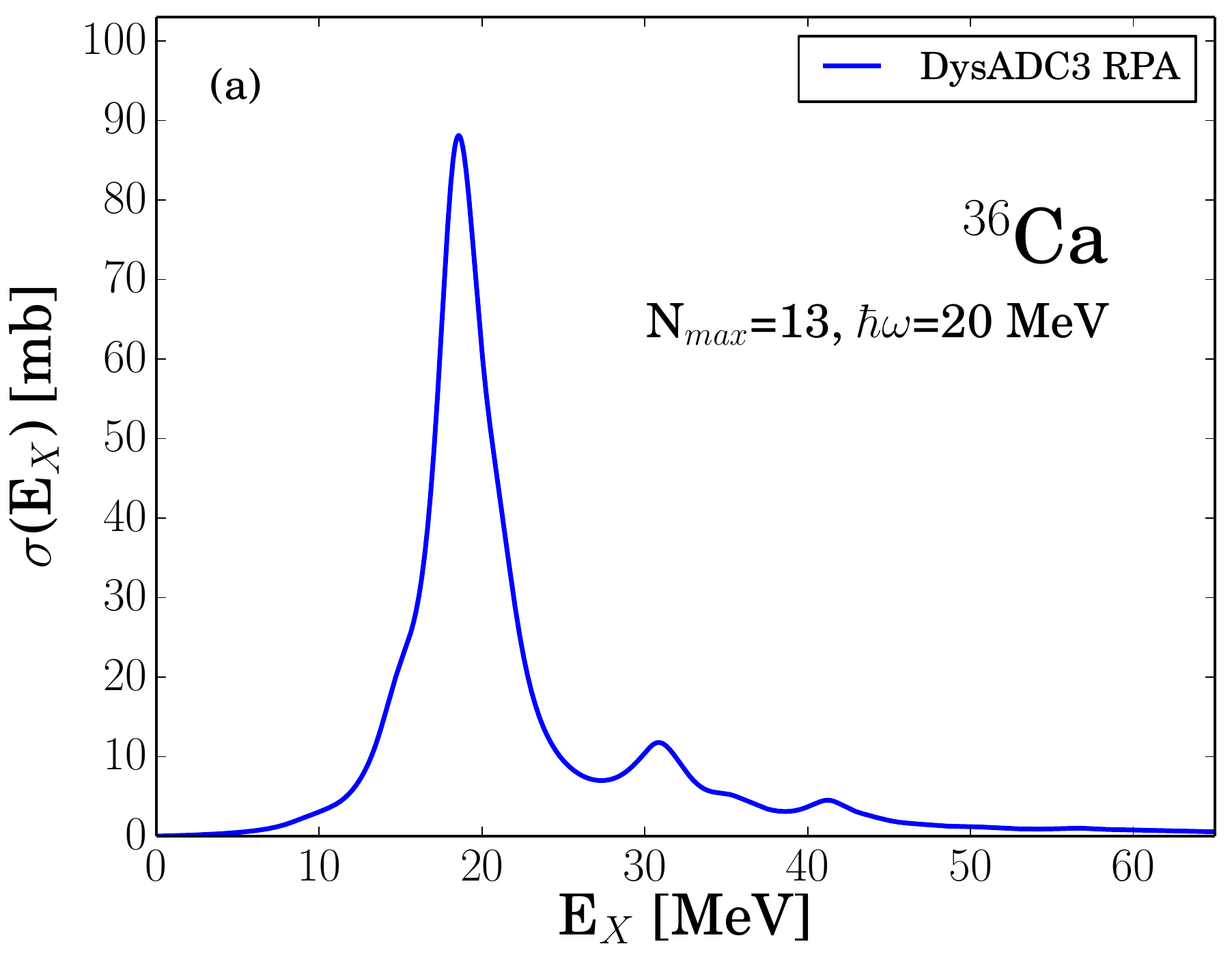}}{\includegraphics[width=\columnwidth,keepaspectratio,clip=true,trim=0.0cm 0.0cm 0.0cm 0.05cm]{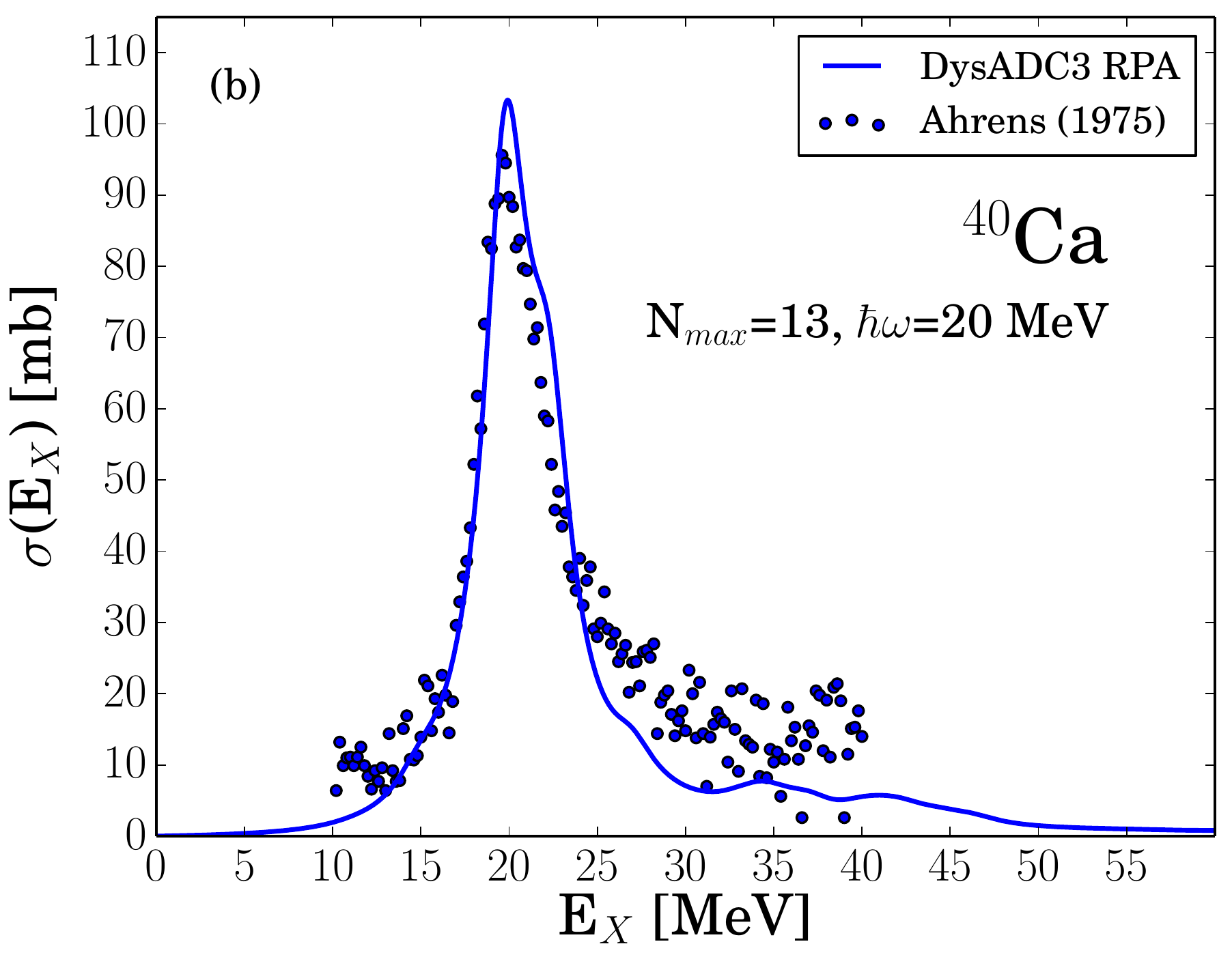}}
{\includegraphics[width=\columnwidth,keepaspectratio]{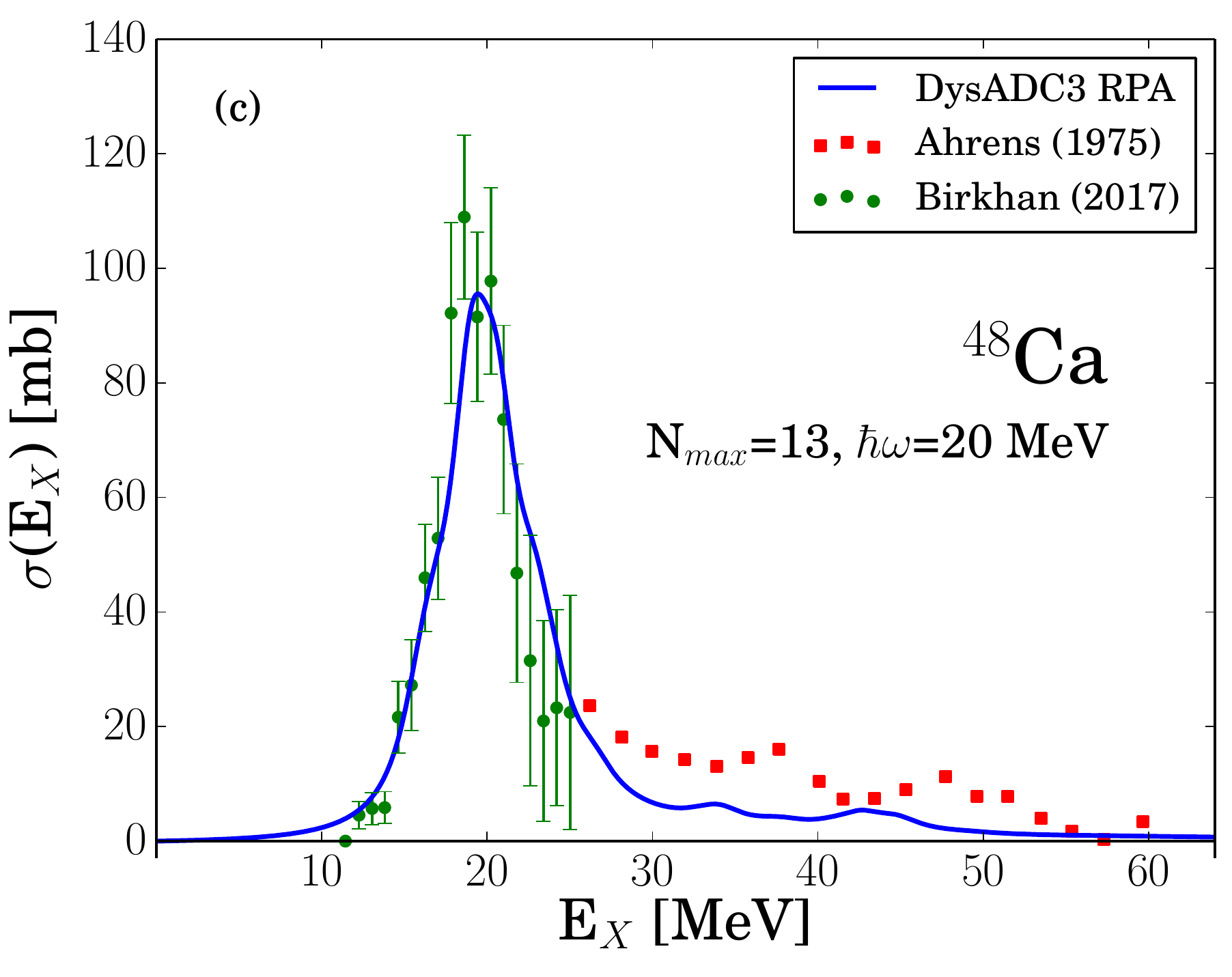}}{\includegraphics[width=\columnwidth,keepaspectratio,clip=true,trim=0.0cm 0.0cm 0.0cm 0.05cm]{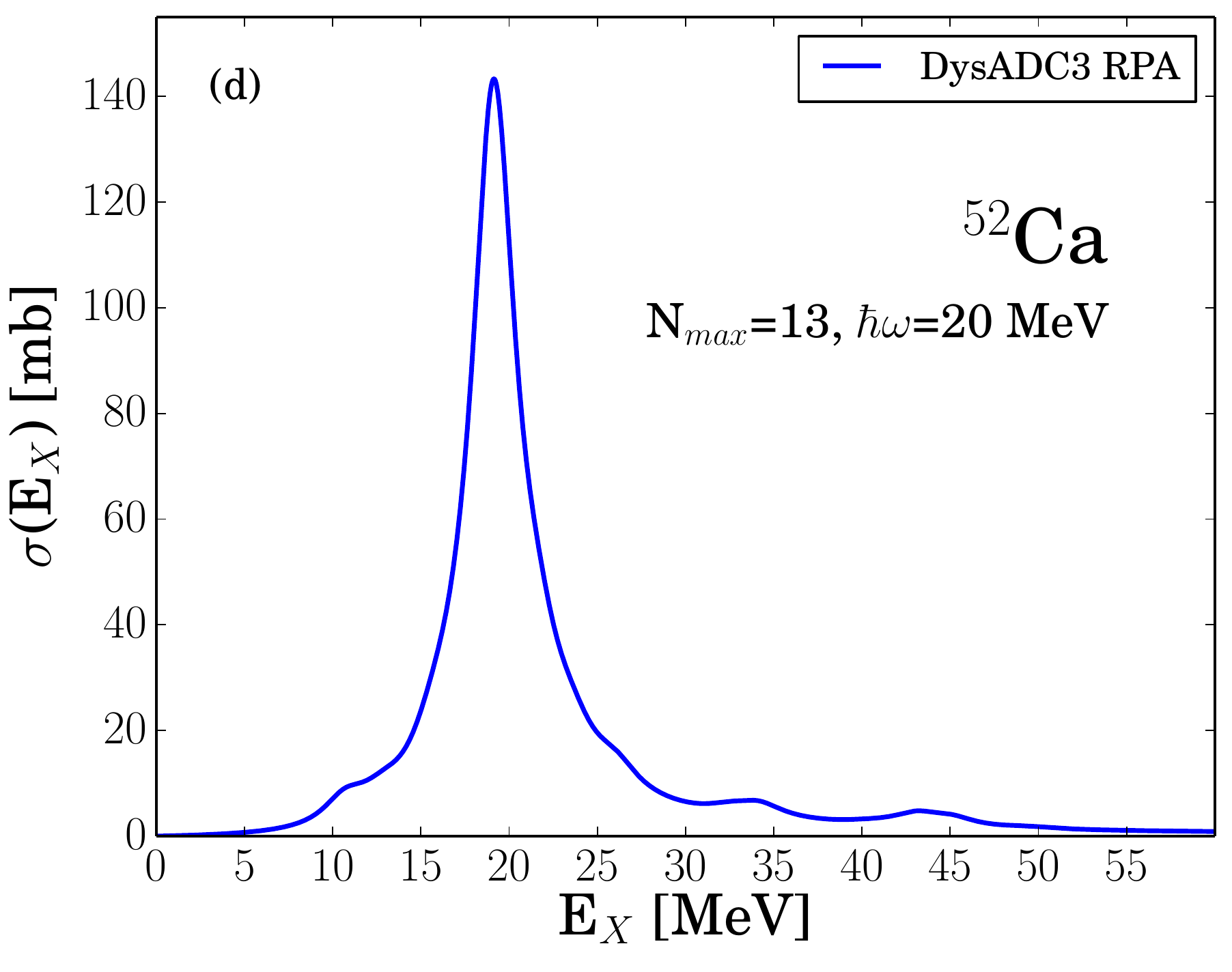}}
{\includegraphics[width=\columnwidth,keepaspectratio]{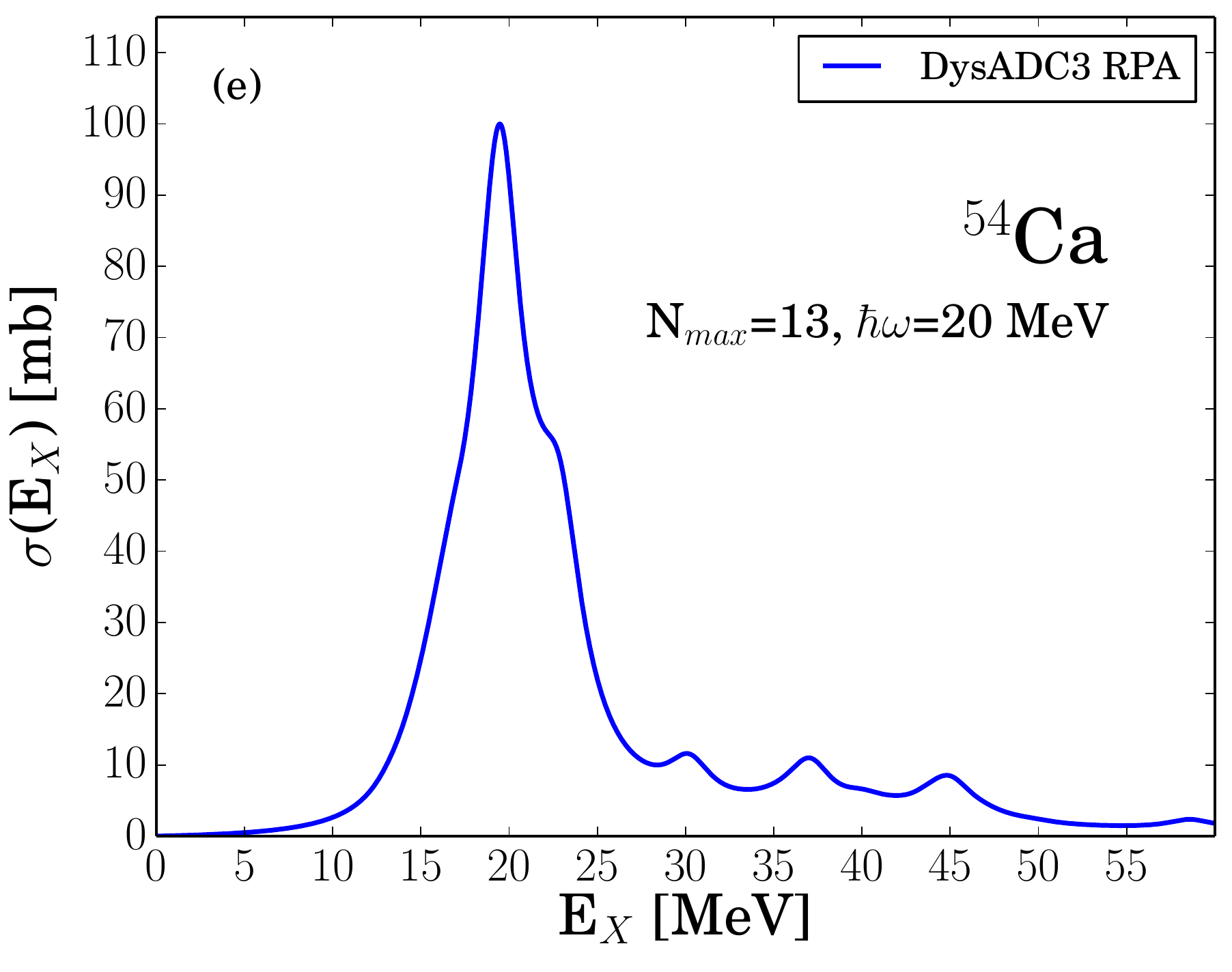}}{\includegraphics[width=\columnwidth,keepaspectratio,clip=true,trim=0.0cm 0.0cm 0.0cm 0.05cm]{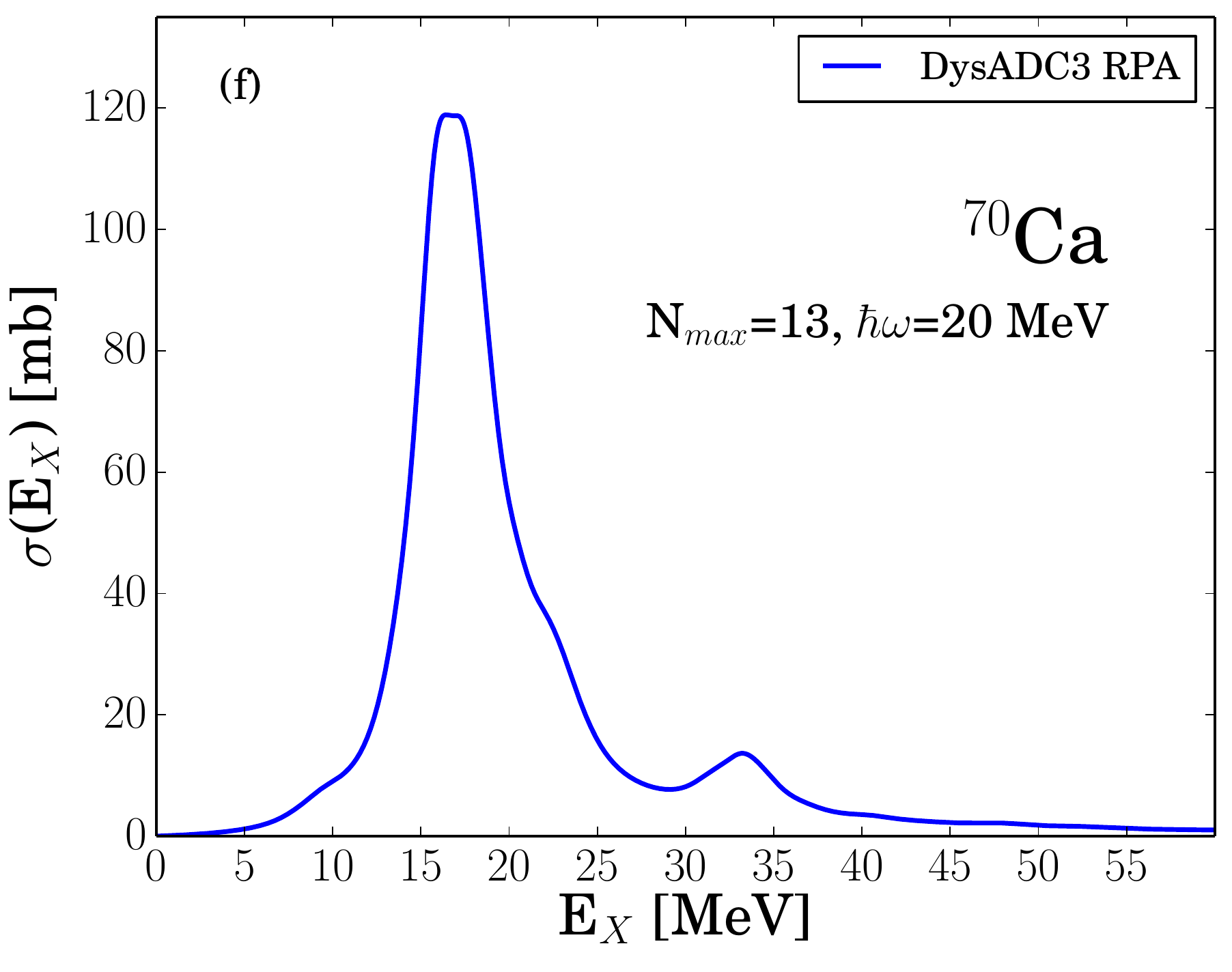}}
\caption{\label{Calcii} 
Same as Fig.~\ref{Oxygens} but for $^{36,40,48,52,54,70}$Ca. Experimental data for $^{40}$Ca in panel~(b) are from Ahrens \textit{et al}~\cite{AHRENS1975479}; experimental data  for  $^{48}$Ca in panel~(c) are from Birkhan \textit{et al}~\cite{PhysRevLett.118.252501} (green circles) and from Ahrens \textit{et al}~\cite{AHRENS1975479} rescaled according to Ref.~\cite{PhysRevLett.118.252501}.
}
\end{figure*}

We have performed calculations of the dipole response and polarizability for 6 Calcium isotopes, from the neutron-deficient $^{36}$Ca to the neutron-rich $^{70}$Ca. The resulting photoabsorption cross sections are displayed in Fig.~\ref{Calcii}. Similarly to Oxygen isotopes, we see that the position of the GDR peak decreases smoothly with the mass number A, in accordance with the slow A$^{-\frac{1}{3}}$ empirical trend. For both $^{40}$Ca and $^{48}$Ca in Figs.~\ref{Calcii}b and~\ref{Calcii}c respectively, the profile of the experimental GDR is well reproduced by our calculation, whereas the high-energy tail of the spectra misses some strength, which  is in line to what we have seen for $^{16}$O. 

 For $^{40}$Ca, the experimental data are available from Ahrens \textit{et al}~\cite{AHRENS1975479,AHRENS1985229}, with the corresponding dipole polarizability displayed in Table~\ref{tab:polariz_Calcii}, where the CC-LIT results~\cite{PhysRevLett.118.252501,PhysRevC.98.014324} are also shown. The experimental value \hbox{$ \alpha_{\text{D}}$ = 1.87(3) fm$^3$} obtained in Ref.~\cite{PhysRevLett.118.252501} by combining data from Refs.~\cite{AHRENS1975479,AHRENS1985229}, is slightly underestimated by the SCGF value and overestimated by the more refined CC-LIT calculation.

The experimental determination of the $^{48}$Ca polarizability has been undertaken recently, in connection with the study of its neutron skin and the correlation with the charge radius~\cite{PhysRevLett.118.252501}. The RPA result with self-consistent Green's funtions is in better agreement with experiment than the CC-LIT calculation. The latter includes $3p3h$ excitations for the ground state and doublets for the excited states, complemented by a consistent treatment of the similarity transformed dipole operator~\cite{PhysRevC.98.014324}. A comparable extension of the SCGF many-body method, going beyond the RPA, would be required to verify the pattern of convergence of these observables with respect to the many-body truncations in our calculations.

\begin{table}[b]
\caption{\label{tab:polariz_Calcii} $^{40}$Ca and $^{48}$Ca isovector dipole polarizabilities $ \alpha_{\text{D}}$  of Eq.~(\ref{polariz}) compared with those calculated with the CC-LIT method in Refs.~\cite{PhysRevC.94.034317, PhysRevLett.118.252501,PhysRevC.98.014324}  and those extracted from the experimental spectra of Ref.~\cite{AHRENS1975479,AHRENS1985229} for $^{40}$Ca
and of Ref.~\cite{PhysRevLett.118.252501} for $^{48}$Ca. }
\begin{ruledtabular}
\begin{tabular}{c||c|c|c}
Nucleus &  SCGF & CC-LIT  & Exp                \\
\hline
$^{40}$Ca  &        1.79        fm$^3$          &         2.23(3)       fm$^3$             &              1.87(3)        fm$^3$              \\
\hline
$^{48}$Ca  &        2.06       fm$^3$          &         2.25(8)       fm$^3$              &              2.07(22)     fm$^3$              \\
\end{tabular}
\end{ruledtabular}
\end{table}

\subsection{$^{68}$Ni}
\begin{figure}
{\includegraphics[width=\columnwidth,keepaspectratio]{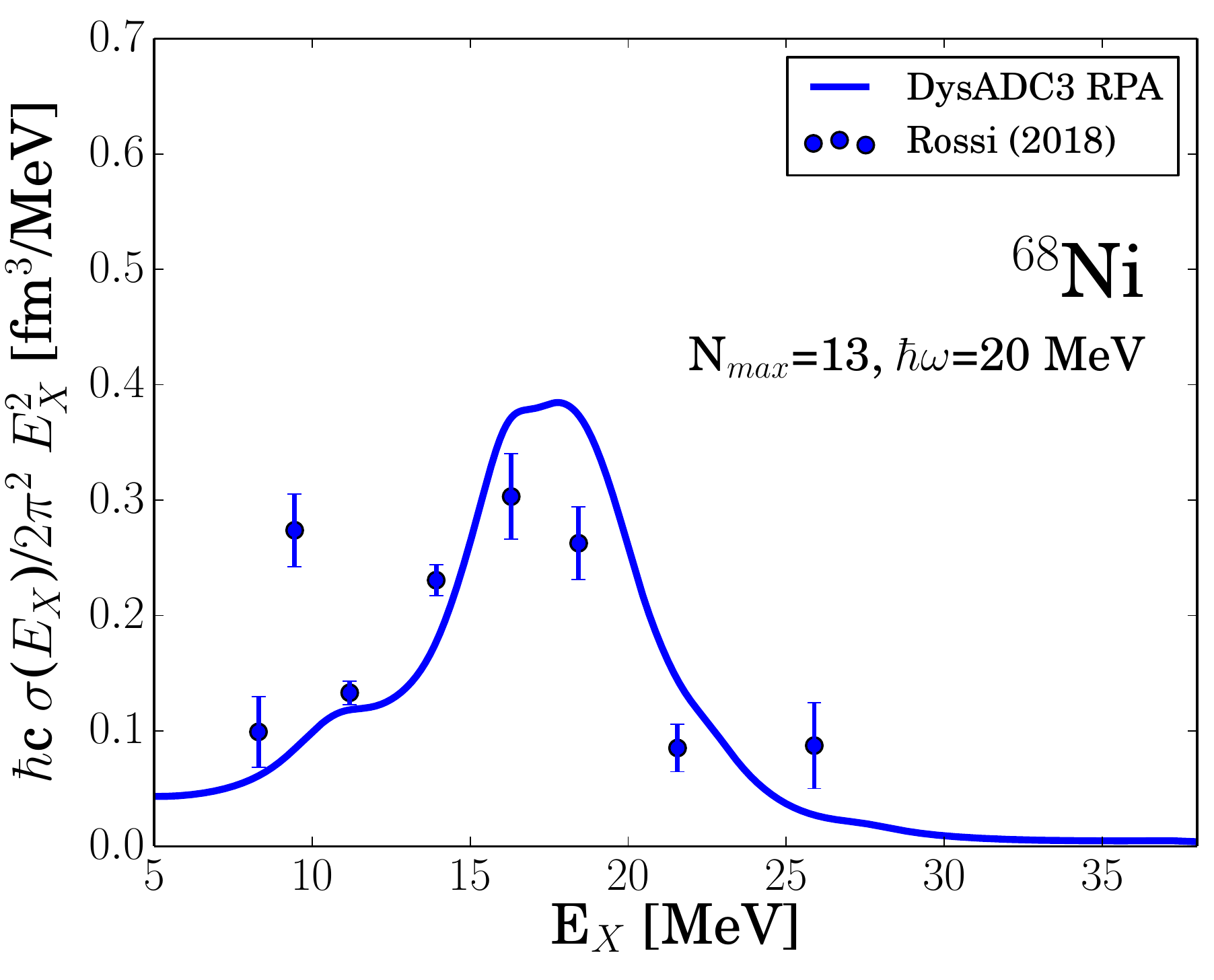}} \\ {\includegraphics[width=\columnwidth,keepaspectratio,clip=true,trim=0.0cm 0.0cm 0.0cm 0.05cm]{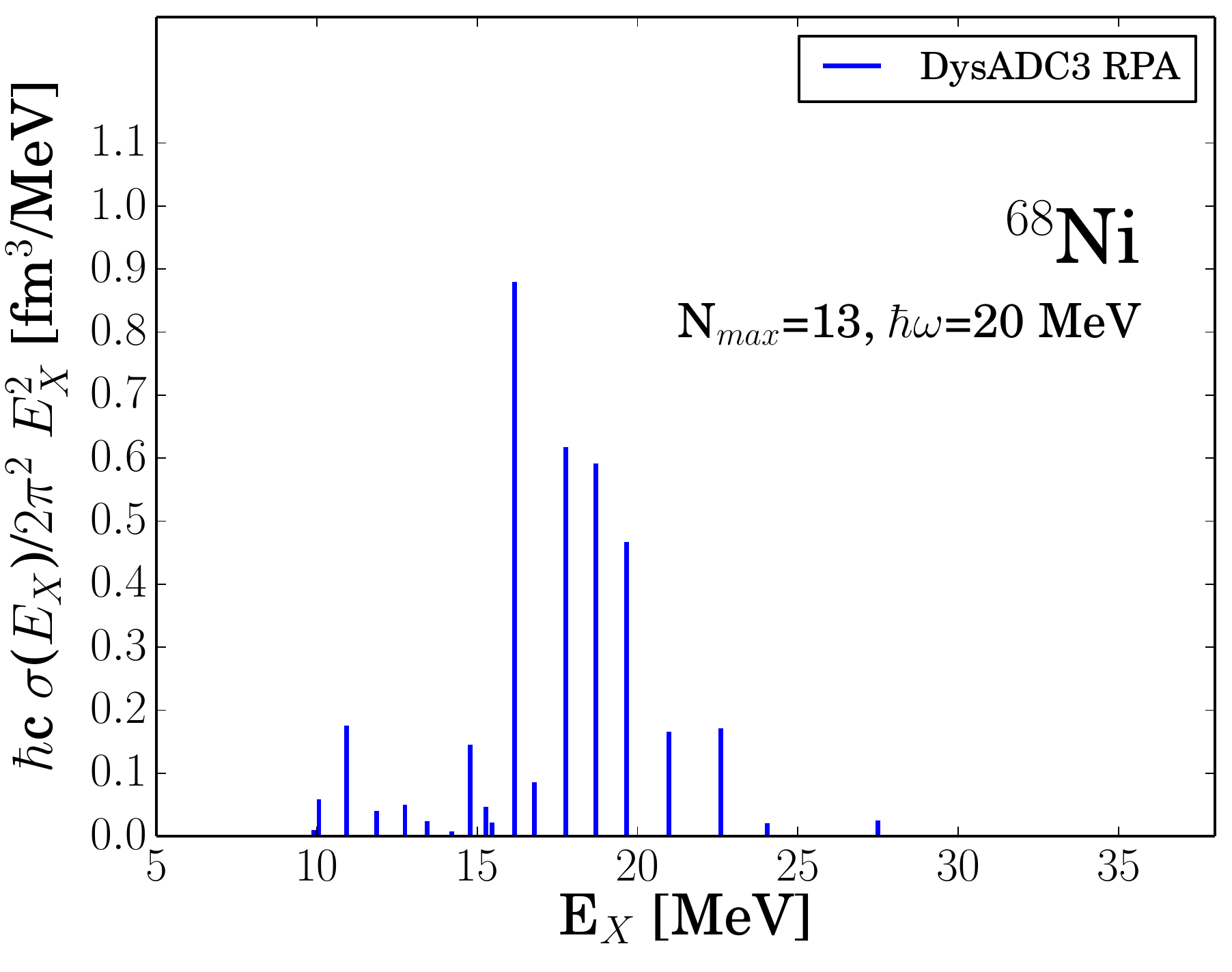}} 
  \caption{  \label{68_Ni}  Isovector dipole response for $^{68}$Ni  computed using a $g^{\text{OpRS}}_{MF}(\omega)$ reference from  Dyson-ADC(3). The lower (upper) panel shows the discrete (convoluted) spectrum obtained from DRPA.  The convolution uses a Lorentzian width \hbox{$\Gamma$ = 3.0 MeV.} Experimental data are  from Rossi \textit{et al}~\cite{PhysRevLett.111.242503}.}
\end{figure}

 The isovector dipole response in the neutron-rich $^{68}$Ni has been recently measured and the corresponding dipole polarizability extracted by Rossi   \textit{et al}~\cite{PhysRevLett.111.242503}. The experimental data are shown in Fig.~\ref{68_Ni} and compared with the computed SCGF curve.
  The few experimental points at $\sim$9.5~MeV  and around $\sim$17~MeV excitation energies are interpreted as Pygmy and Giant Dipole Resonances, respectively. We refer to Table~\ref{tab:Nickel_peaks} for a comparison with the closest peaks in the computed discrete  RPA spectrum, which is also displayed in Fig.~\ref{68_Ni}. In particular, the computed strength at low-energy is fragmented in two principal peaks at 10.68 MeV and 10.92 MeV, located at higher energy than the experimental PDR. For the GDR, Table~\ref{tab:Nickel_peaks} reports the centroid  calculated from the DRPA  response around the main peak after the Lorentzian folding.
 
The $ \alpha_{\text{D}}$ computed by integrating the DRPA spectrum is in agreement with the  experiment, also reported in Table~\ref{tab:Nickel_peaks}. The 3.88(31) fm$^3$ value is obtained by including corrections from a theoretical extrapolation of the low-energy and high-energy parts of the spectrum~\cite{PhysRevC.92.064304},  which were not accessible in the experiment of Rossi \textit{et al}~\cite{PhysRevLett.111.242503}. Both the discrete peaks and the convoluted response in Fig.~\ref{68_Ni} confirm that the computed  spectrum is somehow shifted towards higher energy as compared to the experimental excitation energies. The strength of the PDR is also underestimated.

The lack of strength in the low-energy part of the spectrum could point to insufficient constraints on the isospin-violating contact terms  of the NNLO$_{\text{sat}}$ interaction, giving a too soft symmetry energy. This is also found in calculations of the infinite nucleonic matter within the microscopic Brueckner-Hartree-Fock~\cite{ refId0} and SCGF~\footnote{J. Xu, A. Carbone,  Z. Zhang, and C. Ming Ko, arXiv:1904.09669 [nucl-th]} approaches. However, the limitations of the many-body truncation in the present RPA  scheme prevents us from drawing firm conclusions on the interaction. The correlation between the slope of the symmetry energy and the strength relative to the PDR in $^{68}$Ni  has been verified by using different RPA phenomenological models~\cite{PhysRevC.81.041301}.
When varying the truncation of the model space in our simulations, from small spaces up to convergence, we find that the polarizability of this nucleus is strongly correlated to its radius.

\begin{table}[b]
\caption{\label{tab:Nickel_peaks} Experimental excitation energies of PDR and GDR, and dipole polarizability  in $^{68}$Ni from Rossi \textit{et al}~\cite{PhysRevLett.111.242503}, compared with those calculated with the SCGF method at ADC(3)-DRPA level (see text for details).}
\begin{ruledtabular}
\begin{tabular}{c||c|c}
&  SCGF & Exp                 \\
\hline
PDR [MeV]  &         \begin{tabular}{@{}c@{}}10.68 \\ 10.92 \end{tabular}             &       9.55(17)                             \\
\hline
GDR [MeV]  &           18.1         &         17.1(2)                     \\
\hline
$\alpha_D$ [fm$^3$]  &    \raisebox{\dimexpr-\height + 2.7ex\relax}{3.60}    &       \begin{tabular}{@{}c@{}}3.40(23) \\  3.88(31)  \end{tabular}                         \\
\end{tabular}
\end{ruledtabular}
\end{table}

 \section{\label{other_OpRS}Different reduction of the dressed propagator} 
The procedure for reducing  the fully dressed propagator into a simpler OpRS one  is not unique. Different definitions of the constraining moments can be used, as in Eqs.~(\ref{eq:Mi}) and~(\ref{eq:Mi_bis}). Moreover, propagators  $g_{\alpha\beta}^{\rm{OpRS}}(\omega)$
with different numbers of quasiparticles and quasiholes poles are possible according to the number of moments considered.
In general, the strategy of constraining the lower moments through Eq.~\eqref{moments} is very effective and it works similarly to Krylov subspace projection techniques to induce a fast convergence of the spectroscopic response spectrum~\cite{Soma2014}. As a result, several fundamental observables and physical quantities that are encoded in the fully dressed propagator are retained already when a few moments are conserved. Nevertheless, even with large-scale computational technique it is normally possible to handle only the smallest OpRs propagators. It is therefore interesting to investigate by how much this truncation affects the DRPA computed quantities.
Even more interesting is the need to ascertain the effect of fragmentation, beyond the $g^{\text{OpRS}}_{MF}(\omega)$: as discussed in Sec.~\ref{DRPA_sec}, the fragmented strength in the solution of Eq.~\eqref{eq:Dy} results from admixtures of $2p1h$ and $2h1p$ states. These can couple in the DRPA equations to generate the redistribution of strength at high energies without explicitly including configurations beyond $ph$.  While the above information is washed out of a mean-field propagator,  some fragmentation is already present even in the lowest $\widetilde{g}^{\rm{OpRS}}_{p=0,1,2,\ldots}(\omega)$ reference propagators when the moments~(\ref{eq:Mi_bis}) are constrained.

\begin{figure}[t]
  \centering
    \subfloat{\includegraphics[scale=0.45]{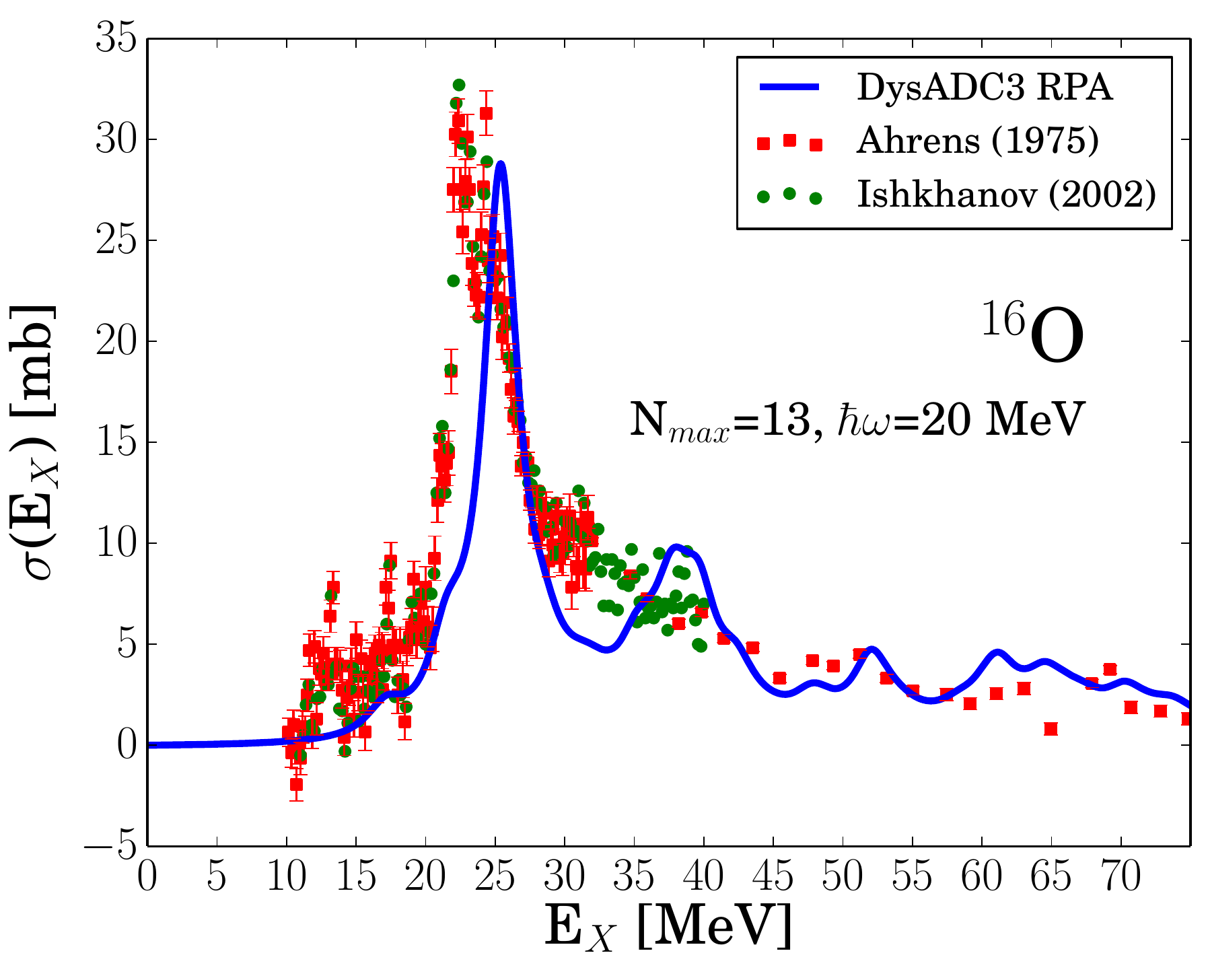}}
  \caption{Photoabsorption cross sections of $^{16}$O computed with  $\widetilde{g}^{\rm{OpRS}}_{p\leqslant 1}(\omega)$. The computed DRPA spectrum is convoluted  with a Lorentzian width of $\Gamma$ = 3.0 MeV.  Experimental data are from Ahrens \textit{et al}~\cite{AHRENS1975479} (red squares) and from Ishkhanov \textit{et al}~\cite{ISHKHANOV2002} (green circles).}
  \label{16O_7poles}
\end{figure}

\begin{figure}[t]
  \centering
    \subfloat{\includegraphics[scale=0.45]{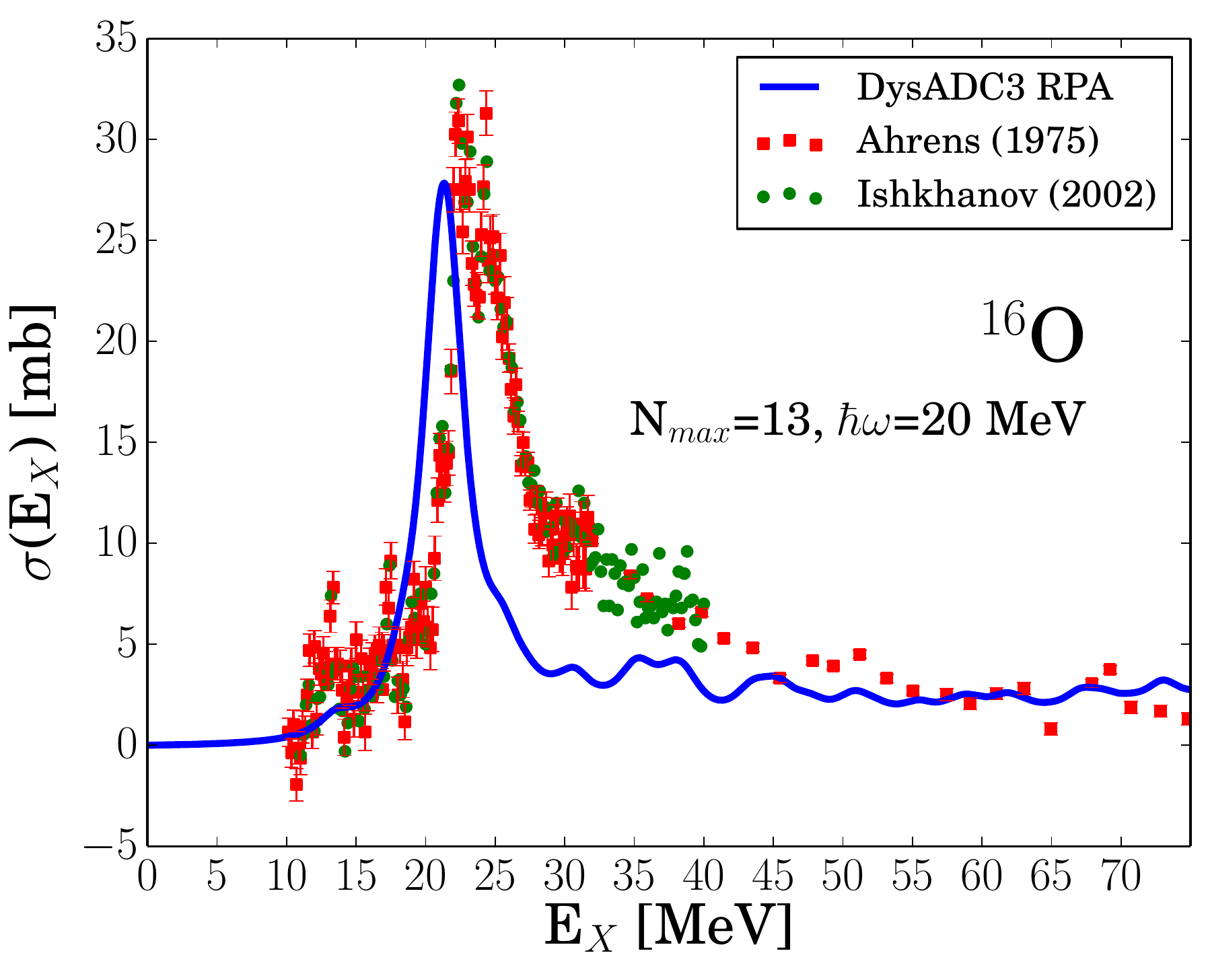}}
  \caption{Same as Fig.~\ref{16O_7poles} but with $\widetilde{g}^{\rm{OpRS}}_{p\leqslant3}(\omega)$.}
  \label{16O_14poles}
\end{figure}

\begin{figure}[t]
  \centering
    \subfloat{\includegraphics[scale=0.43]{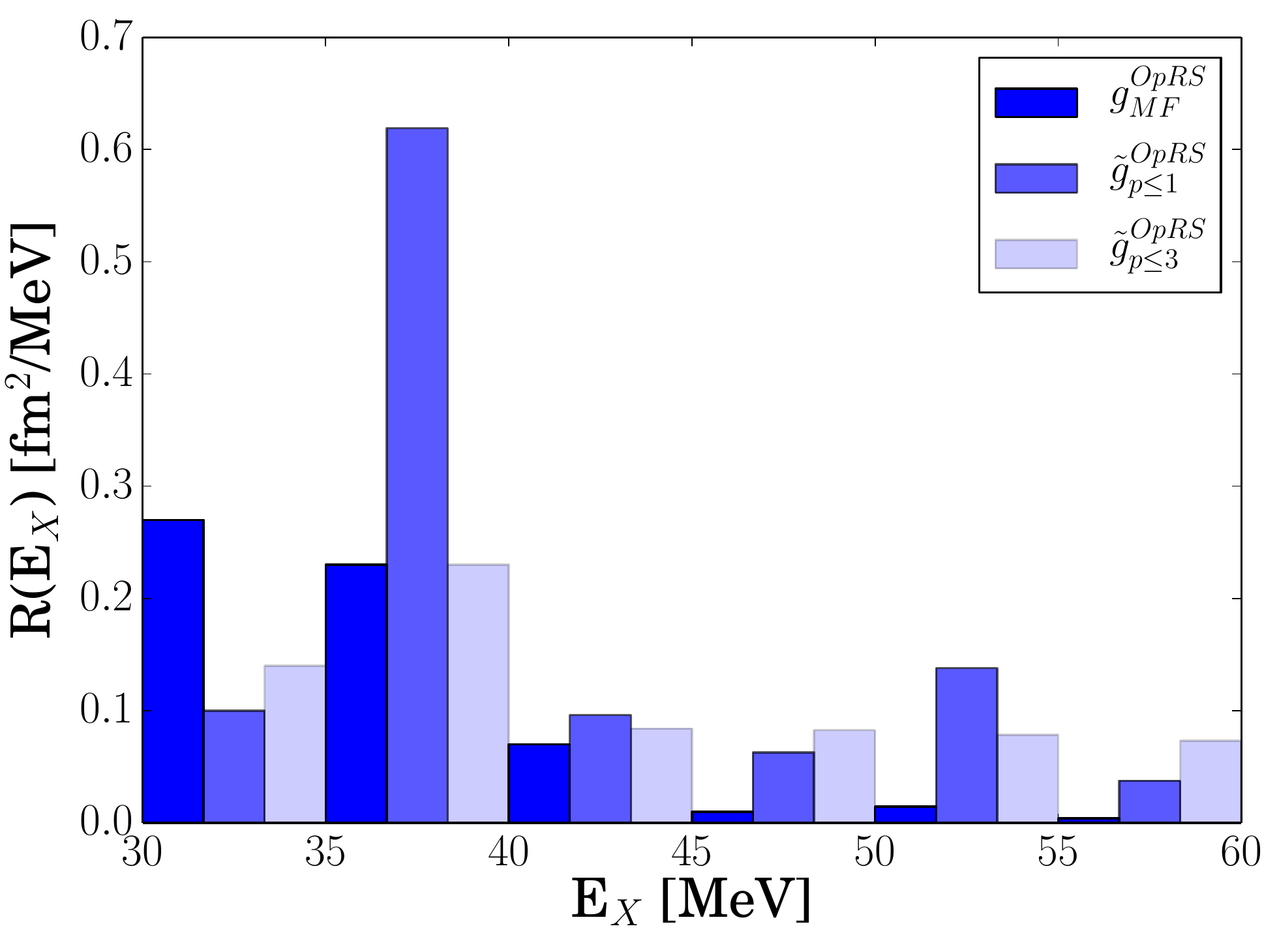}}
  \caption{Integrated E1 strength of $^{16}$O for 5 MeV energy bins in the region of the spectrum above the GDR. The different shaded areas correspond to the mean-field $g^{\rm{OpRS}}_{MF}(\omega)$ (shown in Fig.~\ref{Oxygens}b), the $\widetilde{g}^{\rm{OpRS}}_{p\leqslant 1}(\omega)$ (Fig.~\ref{16O_7poles}) and the $\widetilde{g}^{\rm{OpRS}}_{p\leqslant 3}(\omega)$  (Fig.~\ref{16O_14poles}).}
  \label{16O_strength}
\end{figure}

To investigate these effects, we compare the photoabsorption cross section of $^{16}$O predicted from the mean-field type reference $g^{\rm{OpRS}}_{MF}(\omega)$, which is shown in Fig.~\ref{Oxygens}b, to the DRPA responses resulting from $\widetilde{g}^{\rm{OpRS}}_{p\leqslant 1}(\omega)$  and $\widetilde{g}^{\rm{OpRS}}_{p\leqslant 3}(\omega)$, displayed respectively in Figs.~\ref{16O_7poles} and~\ref{16O_14poles}.
 As expected, the denser spectroscopic fragmentation results in an enhancement of the large energy tail in the excitation spectra. 
 This is better illustrated in Fig.~\ref{16O_strength} where we collected the dipole strength in 5~MeV bins for energies above the GDR. The strength for the $\widetilde{g}^{\rm{OpRS}}_{p\leqslant 1}(\omega)$  and $\widetilde{g}^{\rm{OpRS}}_{p\leqslant3}(\omega)$ becomes important in this energy region compared to the effective propagator of the mean-field type. 
Sec.~\ref{DRPA_sec} also pointed out that the DRPA implies missing diagrams and Pauli violations already among $2p2h$ intermediate states. This issue does not appear to have strong implications in the high energy tail, where several particle and hole fragments are mixed and distributed over a large energy range. However, we find that the centroid energy of the GDR changes notably when introducing more fragmentation. This indicates that full DRPA is better reliable at large energies but becomes unstable in the giant resonance region.   To overcome this limitation, one should implement an extension of the RPA as in Ref.~\cite{PhysRevC.68.014311}, where the polarization diagrams accounting for the coupling with two-phonon contributions are explicitly included.

Finally, we have found that the dipole polarizability $ \alpha_{\text{D}}$ computed with the different OpRS propagators differ by less than 5\%. 
 
\section{\label{concl}Conclusions}
The results discussed in Secs.~\ref{sec:REs} and~\ref{other_OpRS}  show that the SCGF formalism is capable to grasp the main features of the E1 response for the computed nuclei. In particular, the low-energy part of the excitation spectrum up to the GDR is very well reproduced for  $^{16,22}$O and $^{40,48}$Ca, and compares well with  the dipole polarizability values, which involve an integration over the whole experimental spectrum. For the $^{68}$Ni, the energies of both PDR and GDR are slightly overestimated, and we compute a lower strength than the experiment, especially in the PDR.

Despite the fact that the employed $ph$-RPA for the two-body propagator is quite rudimentary, the agreement with the experimental data is comparable to the findings of the higher many-body truncation such as from Ref.~\cite{PhysRevC.98.014324}. This fact should be substantiated by a thorough assessment of the SCGF theoretical errors, for which we still miss - at this stage - direct information regarding many-body truncations beyond the explicit $ph$ intermediate states.  Nonetheless, the results  confirm the good saturation properties of the  NNLO$_{\text{sat}}$ interaction as the crucial mechanism at play in the dipole response, while the missing PDR strength in $^{68}$Ni could be an hint that the isospin dependence of the interaction is not sufficiently constrained~\cite{refId0}. In general, we can conclude that the present DRPA to the Bethe-Salpeter equation performs fairly well for the observables related to the E1 response, where the nuclear correlations included in higher-order diagrams of the polarization propagator could be not so significant.
However, the treatment of the polarization propagator beyond the RPA is expected to be important for higher multipolarities of the electromagnetic response and for the weak processes, requiring an extension of the present implementation.  

The procedure for reducing the fully correlated single-particle propagator to a simpler reference state (referred to as OpRS)  has been tested, and we have concluded that the integrated spectrum has a mild dependence, with a 5\% difference in the dipole polarizability values for different choices of the $g^{\rm{OpRS}}(\omega)$. On the contrary, the dependence of the GDR centroid and of the  response profile in the higher-energy part of the spectrum  points to the limitations of the DRPA, in particular to the incomplete description of interactions among  $2p2h$ and $3p3h$  intermediate state configurations.

The isotopes studied so far are closed-subshell  nuclei, due to the present implementation for the computation of the E1 response being limited to the particle-number conserving version of the SCGF approach. To compute the E1 response for open shell nuclei, an extension of the Gorkov formalism~\cite{Soma2011} to include the two-body propagator is required. The first step of this extension will be equivalent to a microscopic quasiparticle dressed RPA approach.

\begin{acknowledgments}
The authors thank P.  Navr\'atil for providing the interaction matrix elements of the NNLO$_{\text{sat}}$ interaction. Useful discussions with S.\@ Bacca, R.\@  F.\@ Garcia\@ Ruiz and M.\@ Miorelli are also acknowledged.
This research is supported by the United Kingdom Science and Technology Facilities Council (STFC)
under Grants No. ST/P005314/1 and No. ST/L005816/1
 Calculations were performed performed using the DiRAC Data Intensive service at Leicester (funded by the UK BEIS via STFC capital grants ST/K000373/1 and ST/R002363/1 and STFC DiRAC Operations grant ST/R001014/1).
\end{acknowledgments}

\bibliography{RPA_paper_biblio}

\end{document}